%% file: paper.tex
\documentclass[fleqn,usenatbib]{mnras}

\usepackage{newtxtext,newtxmath}

\usepackage[T1]{fontenc}

\DeclareRobustCommand{\VAN}[3]{#2}
\let\VANthebibliography\thebibliography
\def\thebibliography{\DeclareRobustCommand{\VAN}[3]{##3}\VANthebibliography}

\usepackage{graphicx}	
\usepackage{amsmath}	
\usepackage{bm}

\usepackage{neuralnetwork}
\usepackage{tikz}
\usepackage{pgfplots}

\makeatletter
\usepackage{xpatch}
\xpatchcmd{\linklayers}{\nn@lastnode}{\lastnode}{}{}
\xpatchcmd{\linklayers}{\nn@thisnode}{\thisnode}{}{}
\makeatother
\input{nn_customization}

\usepackage{xspace}

\newcommand{\teff}{\ensuremath{\rm{T}_{\mathrm{eff}}}\xspace}
\newcommand{\logg}{\rm{log\ g}\xspace}

\newcommand{\feh}{\rm{[Fe/H]}\xspace}
\newcommand{\hem}{\rm{[He/M]}\xspace}

\newcommand{\msun}{\rm{M_{\sun}}\xspace}
\newcommand{\lsun}{\rm{L_{\sun}}\xspace}

\usepackage{makecell}

\usepackage{booktabs}
\usepackage{graphicx}
\usepackage{pdflscape}
\usepackage{longtable}
\usepackage{multirow}

\usepackage{xcolor, ulem}
\newcommand\hl{\bgroup\markoverwith{\textcolor{yellow}{\rule[-.5ex]{2pt}{2.5ex}}}\ULon}

\title[Generating theoretical RRab light curves using ANN]{Predicting light curves of RR Lyrae variables using artificial neural network based interpolation of a grid of pulsation models}

\author[N. Kumar et al.]{Nitesh Kumar,$^{1}$\thanks{E-mail: niteshchandra039@gmail.com}
Anupam Bhardwaj,$^{2}$
Harinder P. Singh,$^{1}$
Susmita Das,$^{3}$
Marcella Marconi, $^{2}$
\newauthor
Shashi M. Kanbur, $^{4}$
and Philippe Prugniel $^{5}$
\\
$^{1}$Department of Physics and Astrophysics, University of Delhi, Delhi 110007, India \\ 
$^{2}$ INAF-Osservatorio Astronomico di Capodimonte, Salita Moiariello 16, 80131, Naples, Italy \\
$^{3}$ Konkoly Observatory, Research Centre for Astronomy and Earth Sciences, E\"{o}tv\"{o}s Lor\'and Research Network (ELKH), Konkoly-Thege Mikl\'os \'ut 15-17,\\H-1121, Budapest, Hungary\\ 
$^{4}$ Department of Physics and Earth Science, State University of Newyork at Oswego, Oswego, NY 13126, USA \\
$^{5}$ Universit\'e de Lyon, Universit\'e Lyon 1, 69622 Villeurbanne; CRAL, Observatoire de Lyon, CNRS UMR 5574, 69561 Saint-Genis Laval, France
}

\date{Accepted XXX. Received YYY; in original form ZZZ}

\pubyear{2022}

\begin{document}
\label{firstpage}
\pagerange{\pageref{firstpage}--\pageref{lastpage}}
\maketitle

\begin{abstract}
We present a new technique to generate the light curves of RRab stars in different photometric bands ($I$ and $V$ bands) using Artificial Neural Networks (ANN). A pre-computed grid of models was used to train the ANN, and the architecture was tuned using the $I$ band light curves. The best-performing network was adopted to make the final interpolators in the $I$ and $V$ bands. The trained interpolators were used to predict the light curve of RRab stars in the Magellanic Clouds, and the distances to the LMC and SMC were determined based on the reddening independent Wesenheit index. The estimated distances are in good agreement with the literature. The comparison of the predicted and observed amplitudes, and Fourier amplitude ratios showed good agreement, but the Fourier phase parameters displayed a few discrepancies. To showcase the utility of the interpolators, the light curve of the RRab star EZ Cnc was generated and compared with the observed light curve from the {\it Kepler} mission. The reported distance to EZ Cnc was found to be in excellent agreement with the updated parallax measurement from Gaia EDR3. Our ANN interpolator provides a fast and efficient technique to generate a smooth grid of model light curves for a wide range of physical parameters, which is computationally expensive and time-consuming using stellar pulsation codes.
\end{abstract}




\section{Introduction}
RR Lyrae stars (RRLs) are old ($\geq10$ Gyr), population II stars which are in the core helium-burning evolutionary phase of low-mass stars. They are located at the intersection of the horizontal branch (HB) and the instability strip (IS) in the HR diagram. They are low mass ($\sim 0.5 \msun$), metal-poor stars which pulsate primarily in the fundamental mode (RRab) and in the first overtone (RRc) mode with periods ranging from $ \sim 0.2$ d to $ 1.0 $ d \citep[see e.g.][]{bhardwaj_rr_2022}. Many RRL variables are also classified as double mode pulsators (RRd) because they pulsate simultaneously in the fundamental and first overtone modes \citep{sandage_oosterhoff_1981, jurcsik_overtone_2015, soszynski_ogle_2017}. RRLs exhibit a well defined period-luminosity-metallicity (PLZ) relation especially in the near-infrared bands \citep{longmore_rr_1986, bono_theoretical_2001, catelan_rr_2004, sollima_new_2006, muraveva_new_2015, bhardwaj_rr_2021} and thus constitute as primary standard candles of the cosmic distance ladder. RRLs are also excellent probes to gain deeper insights into the theory of stellar evolution and pulsation \cite[e.g.][]{marconi_new_2015, das_variation_2018, marconi_impact_2018}. RRLs are excellent tracers of the old stellar populations due to their age and have been detected in different galactic \citep{vivas_quest_2006, drake_probing_2013, zinn_silla_2014, pietrukowicz_deciphering_2015} and extra-galactic \citep{moretti_leo_2009, soszynski_optical_2009, fiorentino_rr_2010, cusano_dwarf_2013} environments. They have also been identified in several globular clusters \citep{coppola_distance_2011, di_criscienzo_new_2011, kuehn_rr_2013, kunder_rr_2013}.

Advancements in the theoretical modelling of stellar pulsation have made it possible to generate grids of light curves that correspond to a set of physical parameters of the variables using non-linear 1-D hydrodynamical codes, such as those described by \cite{stellingwerf_convection_1984, bono_nonlinear_1997, bono_classical_1999, marconi_new_2015, de_somma_extended_2020, de_somma_updated_2022}. A grid of nonlinear convective hydrodynamical models was produced by \cite{marconi_new_2015} to study the pulsation properties of RRLs. This grid was created using horizontal-branch (HB) evolutionary models \cite[for more information, see][and its references]{pietrinferni_large_2006} and takes into consideration different chemical compositions, luminosity levels, and stellar masses. The models were calculated using the hydrodynamical codes of \citealt{stellingwerf_convection_1982} (updated later by \citealt{bono_theoretical_1998, bono_classical_1999}) and were based on the numerical and physical assumptions detailed in \cite{bono_theoretical_1998, bono_classical_1999} and \cite{marconi_rr_2003, marconi_period_2011}.

Additionally, the radial stellar pulsations of \cite{smolec_convective_2008} available with the Modules for Experiments in Stellar Astrophysics \citep[MESA][]{paxton_modules_2011, paxton_modules_2013, paxton_modules_2015, paxton_modules_2018, paxton_modules_2019}, can also be used to generate the light curves of radially pulsating variable stars.

The theoretical light curves generated from the available pulsation models have been utilized to investigate pulsation properties, derive PLZ relations, and provide a quantitative comparison with the observed light curves of Cepheid and RR Lyrae variables \citep{bono_rr_2000, keller_large_2002, caputo_bright_2004, marconi_pulsational_2005, marconi_modeling_2007, natale_theoretical_2008, marconi_theoretical_2013, marconi_new_2015, marconi_vmc_2017, bhardwaj_large_2017, das_variation_2018, ragosta_vmc_2019, bellinger_when_2020, das_stellar_2020}.

However, generating a model light curve from the given input parameters is still a computationally expensive problem, since it involves solving complex hydrodynamical equations. The processing time has been reduced significantly with modern computational facilities, but theoretical light curves for a fine grid of pulsation models covering the entire parameter space are still not feasible. A smooth grid of pulsation models of RR Lyrae and Cepheids is crucial to predict the physical parameters of these variables either through a model-fitting approach \citep{marconi_theoretical_2013} or using an automated comparison with observed light curves \citep{bellinger_when_2020}. \cite{bellinger_when_2020} obtained a catalogue of physical parameters for observed stars by applying machine-learning methods to the available Cepheid and RR Lyrae models, but the accuracy of physical parameters was limited due to the small number of models in the grid. These authors trained a neural network with light curve structure parameters like $I$ and $V$ band amplitudes, acuteness, skewness, and Fourier coefficients $A_{1}, A_{2}, A_{3}$ (see e.g., \cite{deb_light_2009, bhardwaj_variation_2015}) along with the period as input to predict the physical parameters of the model including mass (M), luminosity (L), radius (R), effective temperature ($\teff$). The theoretical light curves computed by \cite{marconi_new_2015} \& \cite{das_variation_2018} were used for this analysis. They choose the input parameters based on the feature importance study using another ML algorithm, namely Random Forest (RF). The error on the derived parameters is estimated by perturbing the light curve parameters with random noise 100 times and passing them through the ANN.

In this study, we utilized a previously generated grid of models and employed modern automated methods to infer light curves. This was done by training an artificial neural network (ANN) using models from \cite{marconi_new_2015} and \cite{das_variation_2018} to predict light curves based on a set of input parameters. This approach eliminates the need to solve complex time-dependent equations and can generate predictions much more efficiently. 

The paper is organized as follows: The introduction of artificial neural networks and quantitative Fourier analysis is presented in Section~\ref{sec:method}. The input and observational datasets used for network training and comparison are described in Section~\ref{sec:data}. The tuning of the network architecture and training of the final interpolators in both $I$ and $V$ bands is discussed in Section~\ref{sec:ann}. The validity of the trained interpolators is tested by comparing the newly generated model light curves with the ANN-predicted models in Section~\ref{sec:validation_interpolators}. In Section~\ref{sec:interpolator_comparison}, the comparison of Fourier parameters between observed and predicted light curves of LMC and SMC in both $I$ and $V$ bands is discussed, and the distances to these galaxies are estimated. The applications of the trained interpolators are explored in Section~\ref{sec:application}, including a comparison of the observed and predicted $V$ band light curve of a variable star (EZ Cnc) and the determination of its distance in Section~\ref{sec:lc_compare_ez_cnc}. A smooth grid of light curve templates in $I$ and $V$ bands is generated using the ANN interpolators in Section~\ref{sec:new_grid}. The results of the study are summarized in Section~\ref{sec:summary}.

\section{Methodology}\label{sec:method}
The theoretical grid of models can be thought of as being generated via a function $f$ defined as
$$ f (\bm{x}; \bm{w}) = [\text{light-curve}]. $$
where $\bm{x}$ is a combination of physical parameters of the stars, such as M, L, $\teff$, X (hydrogen abundance ratio) and Z (metal abundance ratio), the period and a few other parameters including mixing length, radiative cooling, convective and turbulent flux parameters. Interested readers may refer to \cite{marconi_new_2015} and references therein for a better understanding of the input parameters required for the generation of light curves. Here $\bm{w}$ are fitting parameters. If we assume that the function `$f$' exists and is continuous and differentiable in parameter space: an ANN with one or more hidden layers can reproduce this function. Theoretically, \cite{cybenko_approximation_1989, hornik_multilayer_1989, hornik_approximation_1991} showed that ANN with a non-linearly activated hidden layer, can approximate any continuous function with arbitrary accuracy when provided with enough neurons in the hidden layer. In addition, ANNs are flexible in the choice of architecture, and optimization algorithms, and are simple to implement. They also provide the possibility {\it transfer learning} and therefore is useful in the case of re-training the model with updated data.

\subsection{Artificial Neural Network (ANN)}\label{sec:ann_theory}
\begin{figure}
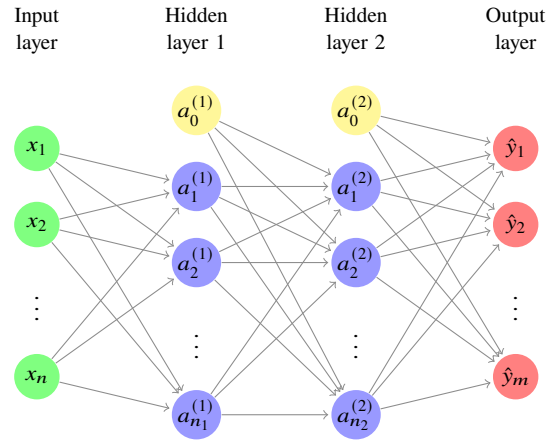

	\begin{center}
		\begin{neuralnetwork}[height=5, style={}, title={}, titlestyle={}, layertitleheight=1.05cm, layerspacing=2.1cm]

			\newcommand{\x}[2]{\ifnum #2=4 $x_n$ \else \small $x_#2$ \fi}
			
			\newcommand{\hfirst}[2]{\ifnum #2=4 $a^{ (1)}_{n_1}$ \else \small $a^{ (1)}_#2$ \fi}
			\newcommand{\hsecond}[2]{\ifnum #2=4 $a^{ (2)}_{n_2}$ \else \small $a^{ (2)}_#2$ \fi}

			\newcommand{\nodetexty}[2]{\ifnum #2=4 $\hat{y}_m$ \else $\hat{y}_#2$ \fi}
			
			\inputlayer[count=4, bias=false, exclude={3}, title=Input\\layer, text=\x]
		
			\hiddenlayer[count=4, bias=true, exclude={3}, title=Hidden\\layer 1, text=\hfirst] \linklayers[not to={3}, not from={3}]

			\hiddenlayer[count=4, bias=true,exclude={3}, title=Hidden\\layer 2, text=\hsecond] \linklayers[not to={3}, not from={3}]
			
			\outputlayer[count=4, exclude={3}, title=Output\\layer, text=\nodetexty] \linklayers[not to={3},not from={3}]
			
			\path (L0-2) -- node{$\vdots$} (L0-4);
			\path (L1-2) -- node{$\vdots$} (L1-4);
			\path (L2-2) -- node{$\vdots$} (L2-4);
			\path (L3-2) -- node{$\vdots$} (L3-4);
		\end{neuralnetwork}
	\caption{A representation of an artificial neural network with $n$ inputs, two hidden layers having $ n_1 $ and $ n_2 $ neurons respectively and an output layer with $ m $ neurons. $ a_{0} $ represents the bias terms in respective layers. The summation and non-linearity nodes are omitted for the sake of clarity.}
	\label{fig:MLP}
	\end{center}
\end{figure}

We employ the simplest neural network which is a feed-forward, fully-connected neural network. The smallest unit of the network, the neuron (or a perceptron) is a mathematical unit that calculates the weighted sum of all the neurons that are previously connected to it and feeds this output to all the neurons in the next layer after applying a non-linear activation function ($\sigma$) (see Fig.~\ref{fig:MLP} ). The value of $i^{th}$ neuron in $k^{th}$ layer is calculated by, 
\begin{equation}
    a_i^{ (k)} = \sigma^{ (k)} \left (\sum_{j=0}^{n_{k}-1} w_{ij}^{ (k-1)} a_{j}^{ (k-1)} \right),\: \text{for} \: 1 \leq i \leq n_{k},
\end{equation}    
with,
\begin{equation}
a_0^{ (k)} = 1 ,  
\end{equation}
here $\sigma^{ (k)}$ is the activation function for the $k^{th}$ layer, which is usually a nonlinear function. A few examples of widely used activation functions are rectified linear unit (or {\it relu}), sigmoid, and hyperbolic tangent ({\it tanh}). The weights of $j^{th}$ neuron that is connected with $i^{th}$ neuron in $k^{th}$ layer are represented by $w_{ij}^{ (k)}$ and are optimised to obtain a minimum (or maximum) value of a certain objective function during the training process. The weights are updated in an iterative process where in each iteration the errors obtained at the output layer are propagated back to the previous layers and thus making the network `better' at every iteration. This algorithm was first conceptualised by ~\citet{Rumelhart86b} and is known as the backpropagation algorithm. This backpropagation algorithm is the backbone of modern stochastic gradient descent (SGD) algorithms \citep[for a review of modern SGD algorithms, interested readers may go through][]{ruder_overview_2016} that optimise the weights of the network based on the quantity that you want to optimise in order to train the network. Fig.~\ref{fig:MLP} represents a schematic of a network with two hidden layers. We have assumed that at for input layer, $k \equiv 0$ and $\bm{a}^{ (0)} \equiv \textbf{x}$ and at the output layer $k \equiv L$ and $\bm{a}^{ (\text{L})} \equiv \hat{\textbf{y}}$ ($\hat{y}$ is the output of the network).

If the predicted and given absolute magnitude value of $i^{th}$ model corresponding to $j^{th}$ phase is $\hat{y}_{ij}$ and $y_{ij}$, the mean square error (MSE) for the $i^{th}$ model is calculated using:
\begin{equation}
    \mathbb{E}_{i} \equiv (\text{MSE})_{i} = \frac{1}{N_s} \sum_{j=1}^{N_s} (y_{ij} - \hat{y}_{ij})^2,
\end{equation} 
where $N_s$ is the total number of magnitude bin values for each model. The average MSE for all models in the grid can be calculated using:
\begin{equation}
\bm{\mathbb{E}} \equiv \text{Avg. MSE} = \frac{1}{N} \sum_{i=1}^{N} (\text{MSE})_{i} = \frac{1}{N} \sum_{i=1}^{N} \left ( \frac{1}{N_s} \sum_{j=1}^{N_s} (y_{ij} - \hat{y}_{ij})^2 \right). 
\end{equation}
where $N$ is the total number of models in the training dataset. 
At every training iteration, for the simplest case of gradient descent learning, the weights are updated according to the following rule:
\begin{equation}
    w_{ij}^{ (k)} = w_{ij}^{ (k)} - \eta \times \frac{\partial \bm{\mathbb{E}}}  {\partial w_{ij}^{ (k)}}
\end{equation}
\noindent where $\eta$ is a scaling factor typically referred to as {\it learning parameter} that determines the size of the gradient descent steps. However, we used a modified version of the gradient descent algorithm known as the adaptive moment optimization algorithm \citep[\texttt{adam},][]{kingma_adam_2014} where the learning parameter is also updated at each iteration to reach the minimum of the objective function efficiently.

\subsection{Fourier parameters}
The analysis of light curves using Fourier analysis has been discussed extensively in literature \citep[][and references therein]{deb_light_2009, bhardwaj_variation_2015, das_variation_2018}. A Fourier series of sines is fitted to the theoretical and observed light curves and the parameters are deduced:
\begin{equation}\label{eq:fourier_sine}
     m = m_{0} + \sum_{k=1}^{N} A_{k} \sin{ (2\pi k \Phi+\phi_k)},
\end{equation}
Here, $A_k$ and $\phi_k$ are Fourier amplitude and phase coefficients, respectively, and $\Phi$ is the pulsation phase that ranges from 0 to 1, which is calculated using the following equation:
\begin{equation}
    \Phi = \left [\frac{t-t_0}{\rm P} \right] -\texttt{Int}\left [\frac{t-t_0}{\rm P} \right], 
\end{equation} 
here, $t_0$ is the epoch of maximum brightness, and P is the pulsation period of the star/model.

$A_k$ and $\phi_k$ are used to calculate Fourier amplitude ratios ($R_{k1}$) and phase differences ($\phi_{k1}$):
\begin{equation}\label{eq:Rk1}
    R_{k1} = \frac{A_k}{A_1},
\end{equation}
\begin{equation}\label{eq:phi_k1}
    \phi_{k1} = \phi_{k} - k\phi_1,
\end{equation}
where, $k > 1$ and $0 \leq \phi_{k1} \leq 2\pi$.

\section{Data}\label{sec:data}
\subsection{Training data}
\begin{figure*}
    \centering
    \includegraphics{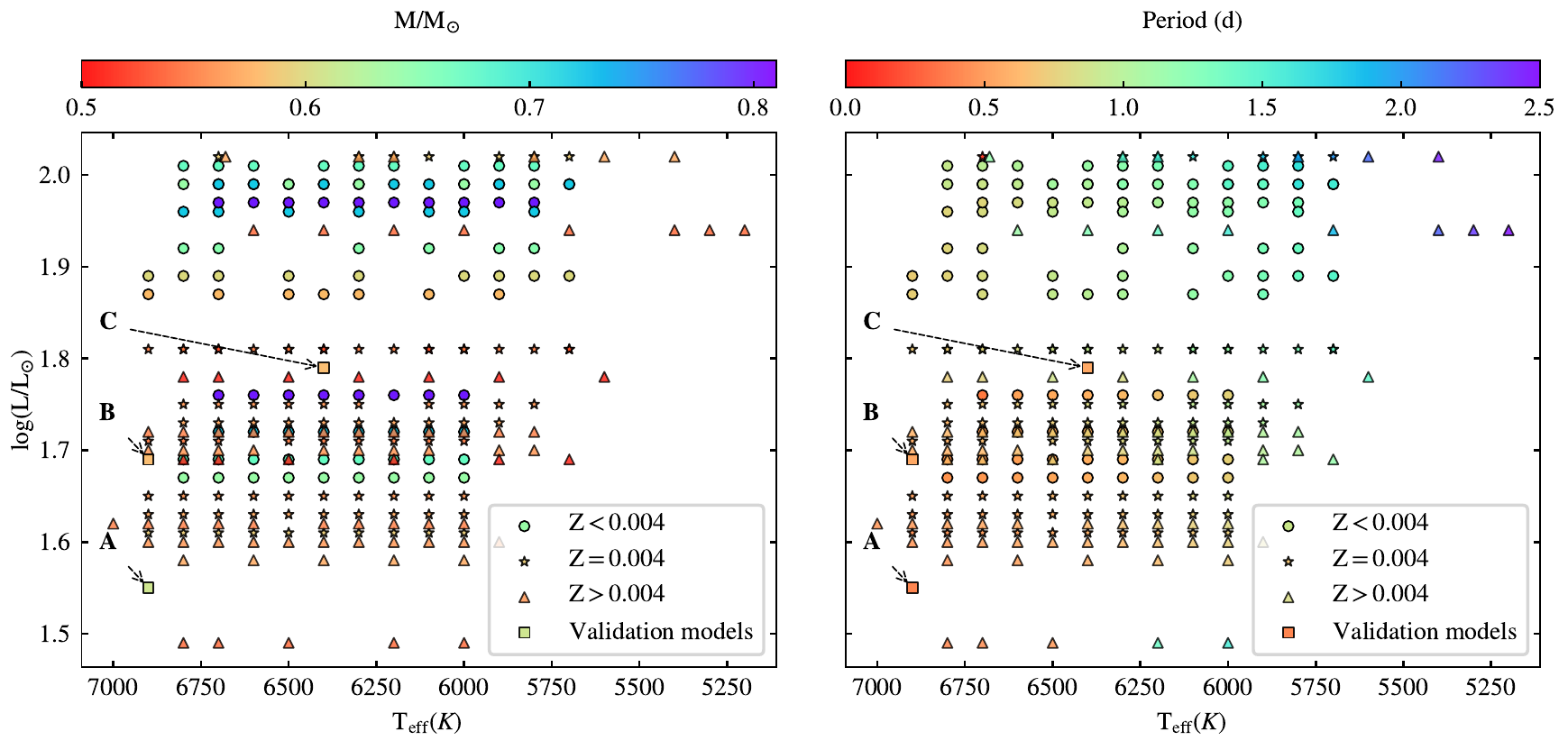}
    \caption{A visual representation of the input dataset of RRLs models described in Table~\ref{tab:input_dataset}. A few additional models were computed in this work for the validation of the trained ANN interpolators and are marked with $\textbf{A}$, $\textbf{B}$, and $\textbf{C}$}.
    \label{fig:input_data}
\end{figure*}

To train the neural network, we adopted the theoretical light curves of fundamental mode RR Lyrae (or RRab) models as described in \citet{das_variation_2018}. We used a total of $274$ RRab light curves corresponding to a grid of physical parameters, out of which $166$ were initially computed by \citet{marconi_new_2015} and the additional $108$ by \citet{das_variation_2018} using the same non-linear, time-dependent convective hydrodynamical models. The model light curves were generated in bolometric magnitudes and they were later transformed into Johnson Cousin photometric bands. A summary of theoretical RRab models is presented in Fig.~\ref{fig:input_data}. The models contain seven distinct chemical compositions ranging from $Z = 0.001$ to $Z = 0.02$, with a primordial He abundance, $Y = 0.245$, and a constant helium-to-metals enrichment ratio, $\hem = 1.4$ to replicate the initial helium abundance of the sun \citep{serenelli_determining_2010}. The pulsation models were constructed with different sets of stellar masses and luminosities, which were fixed according to detailed central He-burning horizontal-branch evolutionary models. The range of Z is broad enough for the comparison with the observed RR Lyrae stars in the satellite galaxies of the Milky Way, the Small Magellanic Cloud (SMC), and the Large Magellanic Cloud (LMC) \citep{clementini_distance_2003}. A few models have pulsation periods greater than one day, thereby including the possibility of evolved RR Lyrae stars in the training data set. Out of the $274$, six overlapping models were removed from the input dataset leaving $268$ unique RRab models to train the ANN. The ANN models are described in detail in Section~\ref{sec:ann}.

\begin{table}
    \centering
    \caption{A summary of 268 Fundamental mode RR Lyrae models.}
    \small
\input{table/parameter_table}
    \label{tab:input_dataset}
\end{table}

\subsection{Observational data used for comparison}\label{sec: obs_data}
We used the observed light curves for a comparison with the ANN predicted light curves. Since the training light curves are in the Johnson Cousin filters bandpass, we need to have the observed light curves in the same filters. We used the $I$ and $V$ band light curves from the IV$^{\rm th}$ phase of the Optical Gravitational Lensing Experiment (OGLE\footnote{\url{http://ogle.astrouw.edu.pl/}}) catalogue of RR Lyrae variables in LMC and SMC \citep{soszynski_ogle_2016}. 

However, the observed magnitudes suffer from interstellar extinction and reddening due to the interaction of light with interstellar dust. We need to account for interstellar extinction and perform extinction corrections in magnitudes at a given wavelength. Using the positions (RA/Dec), we obtain the excess colour values, $E (V - I)$, for RRLs in the LMC and SMC from the reddening maps of \cite{haschke_new_2011}. { We apply extinction corrections using Schlegel's list of conversion factors\footnote{\url{https://dc.zah.uni-heidelberg.de/mcextinct/q/cone/info}}} -  
\begin{equation}\label{eq:Av}
    A_V = 2.4 \times E(V-I) = 3.32 \times E(B-V),
\end{equation}
 and
\begin{equation}\label{eq:Ai}
     A_I = 1.41 \times E(V-I) = 1.94 \times E(B-V).
\end{equation}

\section{ANN Interpolator for RR Lyrae models}\label{sec:ann}
We used the ANN to interpolate the light curve for the input parameters within the grid. We built a feed-forward fully connected neural network and optimised its network architecture. The input layer of this network takes six parameters, which are: 
$$\bm{x} \equiv \left[\frac{\text{M}}{\msun}, \,  \log \left(\frac{\text{L}}{\text{L}_{\sun}}\right), \, \teff, \, \text{X}, \, \text{Z}, \, \log (P) \right],
$$
where M and L are the mass and luminosity of the model. $\teff$ is the effective temperature of the model in Kelvin. X and Z are the hydrogen and metal abundances of the model and the parameter P is the period in days. We have included the period as one of the inputs to facilitate ease for the user who might want to generate a model corresponding to a specific period.

The hydrodynamic models presented in \cite{marconi_new_2015} were generated using the same physical and numerical assumptions employed in earlier works, such as \cite{bono_theoretical_1998, bono_classical_1999} and \cite{marconi_rr_2003, marconi_period_2011}. For instance, the radiative opacities used in their models were taken from the OPAL radiative opacities provided by the Lawrence Livermore National Laboratory \citep{iglesias_updated_1996}, while the molecular opacities were obtained from \cite{alexander_low-temperature_1994}. Since the physical conditions, including the radiative and molecular opacities were kept constant, they were not considered as inputs to the ANN.

Each input parameter has a different numerical range, we need to transform each parameter to have the same numerical range as other input parameters. The input parameters are scaled in such a way that each input parameter has zero mean and unity standard deviation over the whole grid. At the output layer, we provide the corresponding ($I$ or $V$ band) light curve of the model which consists of the absolute magnitudes at a given series of phases. We have a total of $1000$ magnitude values per model corresponding to phase values from $0$ to $1$ in steps of $0.001$. We used a neural network that has one input layer with six input neurons, a few hidden layers (not more than three to keep the network small), and one output layer with $1000$ output neurons containing the absolute magnitude corresponding to phase values between $0$-$1$. We used a linear activation function ($\sigma^{ (L)} \equiv 1 )$ between the last hidden layer and the output layer because we do not want to constrain the output values in any particular range. We have a total of $268$ models, and hence the training matrix at the input layer has [$268 \times 6$] and at the output layer is [$268 \times 1000$].

\subsection{Network Architecture and Hyperparameter optimisation}
To have a generalised network that does not overfit/underfit the given dataset, we need to choose a suitable architecture (the number of hidden layers, the number of neurons in the hidden layer, activation functions) as well as the learning hyperparameters such as the loss optimization algorithm and parameters therein, etc \citep{elsken_neural_2019}. Each individual hyperparameter has a significant role in the training process and a different value of a hyperparameter significantly affects the result of the training. However, there are no explicit `rules' for selecting these attributes in such a way that the ANN model does not become trapped in a local solution. This is a crucial problem in the field of machine learning \citep{guo_novel_2008}. 

The choice of architecture and hyperparameters usually depends on the intuition of the expert and hand tuning. Typically, the trial-and-error approach like grid search and random search \citep{bergstra_random_2012} is used to determine these characteristics. We created a grid of possible hyperparameters by choice and intuition that we gained working with the dataset, which is tabulated in Table~\ref{tab:hyperparameter_combinations}. Out of various possible combinations, we chose a set of $100$ hyperparameter combinations at random. We then trained the network for fixed $1000$ epochs with the L2 norm (MSE) as the objective function using the ``adaptive moment stochastic gradient descent (or \texttt{adam}: \citealt{kingma_adam_2014})'' algorithm with a default batch size of $32$ samples. The network architecture was optimised using the $I$ band light curves. For this procedure, we utilised \texttt{Keras tuner} \citep[][]{omalley_kerastuner_2019}.

\begin{table}
    \centering
    \caption{The hyperparameter search space for the network.}
    \resizebox{\linewidth}{!}{%
    \begin{tabular}{cll}
        \hline 
        S.No. & Name of hyperparameter & Possible values  \\
        \hline 
        1 & No. of hidden layers & [1, 2, 3] \\
        2 & No of neurons in  one hidden layer & [16, 32, 64, 128]   \\
        3 & Optimiser & `\texttt{adam}' \citep{kingma_adam_2014} \\
        4 & Learning rate  & [$10^{-2} - 10^{-4}$](log sampling) \\
        5 & Activation function & [`\texttt{relu}', `\texttt{tanh}'] \\
        6 & Weights initialisation & `\texttt{GlorotUniform}' \\
        \hline
    \end{tabular}
    }
    \label{tab:hyperparameter_combinations}
\end{table}

\begin{table*}
\centering
\caption{The performance of different neural network architectures trained with the different combinations of hyperparameters.}
\input{table/hparams_tune}
\label{tab:hparams_tune}
\end{table*}

\begin{figure}
    \centering
    \includegraphics{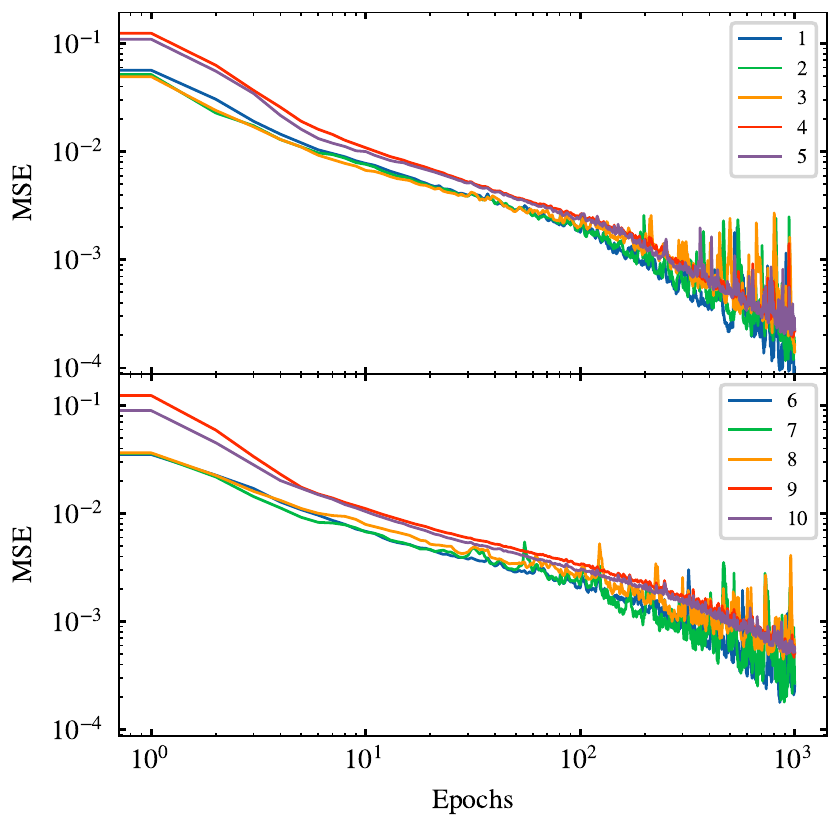}
    \caption{Training curves of the network architectures described in Table~\ref{tab:hparams_tune}. We seek such an architecture that converges to the minimum rapidly in the first 1000 epochs. We find that the network with architecture $1$ reaches the minimum fastest.}
    \label{fig:train_curve_1}
\end{figure} 

We show the result of the top 10 performing network architectures in Table~\ref{tab:hparams_tune}. We observe that a network with $3$ hidden layers with $64, 128, 128$ neurons in successive hidden layers with an initial learning rate$(\eta) \sim 1.8 \times 10^{-3}$, reaches the minimum MSE in $1000$ epochs of learning. Fig.~\ref{fig:train_curve_1} depicts the learning curves for the top-10 network architectures listed in Table~\ref{tab:hparams_tune}.  

\subsection{Training of the I band Interpolator}
\begin{figure*}
    \centering
    \includegraphics{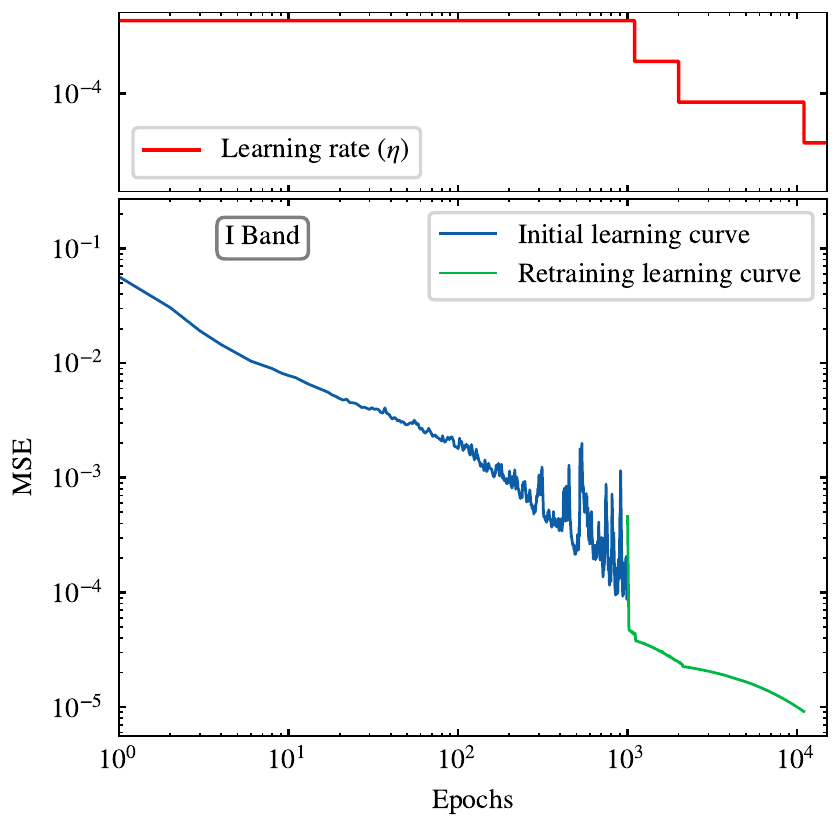}
    \includegraphics{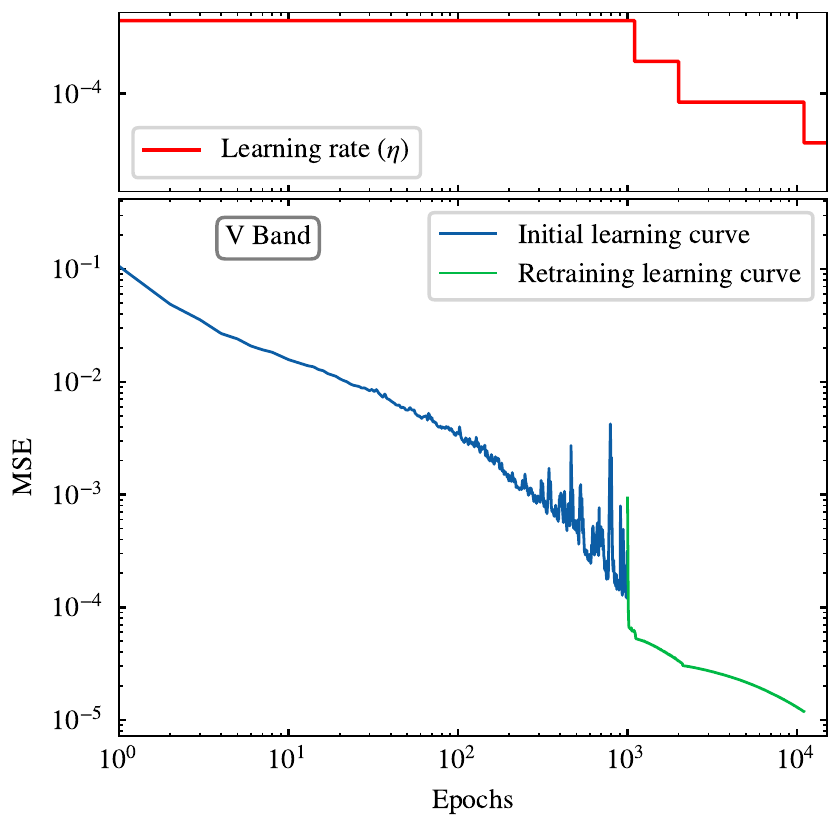}
    \caption{ The learning curve for both the $I$ and $V$ band is shown in the respective blocks along with the learning rate parameter. The blue curve in each plot represents the training curve till the initial 1000 epochs trained with the best-performing network architecture obtained using random search. We then transfer the trained weights and re-train the network with piece-wise decay of the learning rate. The orange curve represents the training curve in this phase.}
    \label{fig:train_curve_2}
\end{figure*}
To construct the final interpolator for the $I$ band, we used the architecture that performed best in the architecture optimization step, i.e., architecture 1. In the previous step, we note that the training loss (MSE) decreases globally at large epochs, but the individual updates are quite big and the loss oscillates rapidly (see Fig.~\ref{fig:train_curve_1}). High learning rates cause such oscillations in the learning curve. The oscillations can be reduced by reducing the learning parameter; however, we don't want to start the training with a low $\eta$ because it impairs the performance of the network (see the comparison between architecture no. 1 and 4: the low learning rate causes the network to return comparatively higher MSE). Hence, we employ `piece-wise decay' of the learning rate to mitigate these oscillations of loss at higher epochs. To do this, we begin with the best-performing network of the previous step and re-train it with a decreasing learning rate. We reduce the current learning rate by dividing it by a constant number (`$\delta$') when the epoch crosses successive powers of $10$. We chose $\delta = 5$, based on the intuition we gained after experimenting with a few values of $\delta$. This guarantees that the loss diminishes gradually as the training epochs increase. We trained this network for $11,000$ epochs and obtained a minimum MSE of $9.07 \times 10^{-6}$. The final learning curve (epoch vs loss curve) is shown in Fig.~\ref{fig:train_curve_2} along with the learning parameter ($\eta$) in the top panel. This final network is adopted as the $I$ band ANN interpolator for fundamental mode RR Lyrae models.

\subsection{Training of the V band Interpolator}
The problem of training the network for interpolating the light curve in a different band is a similar problem that we tackled in the previous step and hence we used the same network architecture that performed best during the $I$ band interpolator training, to train the $V$ band interpolator. We started the training with a neural network with $3$ hidden layers which contain $64, 128, 128$ neurons respectively. We started to train the network with the same `\texttt{adam}' optimization algorithm with the same initial learning rate parameter ($\eta = 1.7941 \times 10^{-3}$). After the initial training for $1000$ epochs, we encounter the same problem of loss oscillations, and hence we treat the learning rate in the same manner as we did in the case of the $I$ band interpolator. The learning curve for the $V$ band interpolator along with the adopted learning rate parameter is shown in the right panel of Fig.~\ref{fig:train_curve_2}.

\subsection{Training statistics for interpolators}\label{sec:testing_interpolators}
We calculated the statistics between the original model light curves and the ANN generated/predicted light curves. We determined the average mean squared error (MSE), mean absolute error (MAE) and the coefficient of determination ($R^{2}$) between the original and predicted magnitude values for a quantitative comparison \citep{steel_principles_1960, draper_applied_1998, glantz_primer_2001}. We have already discussed the MSE in Section ~\ref{sec:ann_theory}. The MAE for one model is defined by:
$$ \text{MAE} = \frac{1}{N_s} \sum_{j=1}^{N_s}  |y_{j} - \hat{y}_{j}|.  $$
and $R^{2}$ is defined by: 
$$ R^{2} = 1 - \frac{ \sum (y_j - \hat{y}_j)^{2} }{\sum (y_j - \Bar{y})^2}. $$
where $y_j$ and $\hat{y}_j$ is the original and predicted magnitude value corresponding to the $j^{th}$ phase and $\Bar{y} = \frac{\Sigma y_j}{N_s}$. $N_s$ is the total number of sample points for the light curve. We quote the average of each quantity over the whole dataset.

The training statistics for both $I$ and $V$ bands are given in Table~\ref{tab:train_stats}. For the $I$ band interpolator, we achieve a minimum MSE of $\sim 9.076 \times 10^{-6}$, with corresponding MAE of $ \sim 2.162 \times 10^{-3}$ and $R^{2} \sim 0.9987$. For the $V$ band interpolator, we achieve a minimum MSE of $1.175 \times 10^{-5}$, with corresponding MAE of $ \sim 2.503 \times 10^{-3}$ and $R^{2} \sim 0.9994$.

\begin{table}
    \centering
    \caption{The statistics of the interpolators on training data.}
    \begin{tabular}{cccccc}
    \hline
         Band & No. of Models & MSE & MAE      & $R^{2}$ \\
     \hline
         I    & 268          & $9.076 \times 10^{-6}$  & $ 2.162 \times 10^{-3}$& $ 0.99879$ \\
         V    & 268          & $1.175 \times 10^{-5}$  & $ 2.503 \times 10^{-3}$& $ 0.99940$ \\
    \hline
    \end{tabular}
    \label{tab:train_stats}
\end{table}

\subsection{Validation of Interpolators}\label{sec:validation_interpolators}

\begin{figure*}
    \centering
    \includegraphics[scale=0.98]{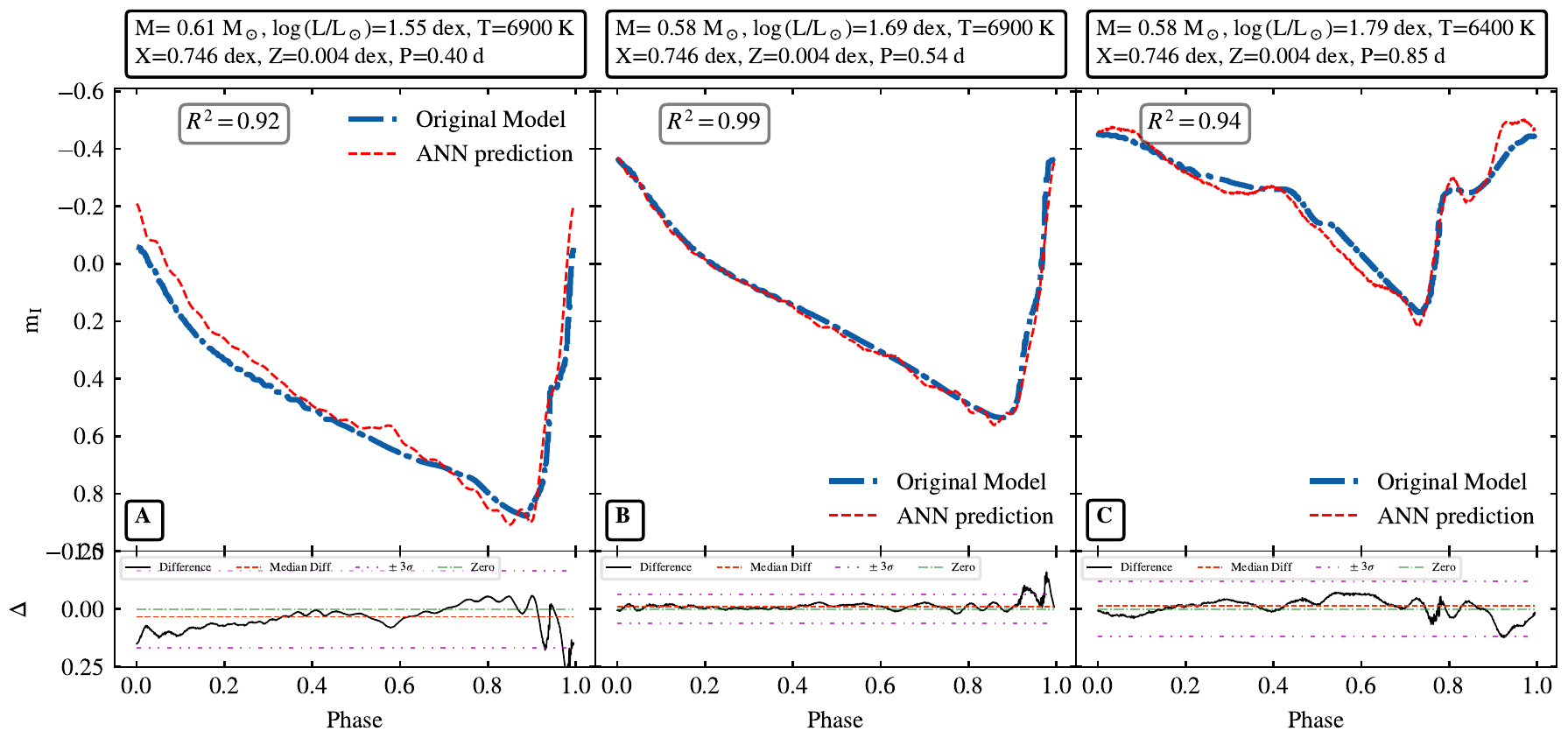}
    \caption{The comparison of the original $I$ band light curve, represented by the blue background line, and the ANN reconstructed light curve, shown as the foreground red line, is displayed for three models. These light curves were generated using the same hydrodynamical code for validation purposes. The input parameters for each light curve plot can be found in the upper panel. The difference between the two light curves is depicted in the bottom panel with a black line, with the magenta line representing the $3\sigma$ bounds and the green line indicating the mean deviation between the predicted and actual light curves. The goodness of fit parameter $R^{2}$ is also calculated for each plot and displayed.}
    \label{fig:validation_I}
\end{figure*}

\begin{figure*}
    \centering
    \includegraphics[scale=0.98]{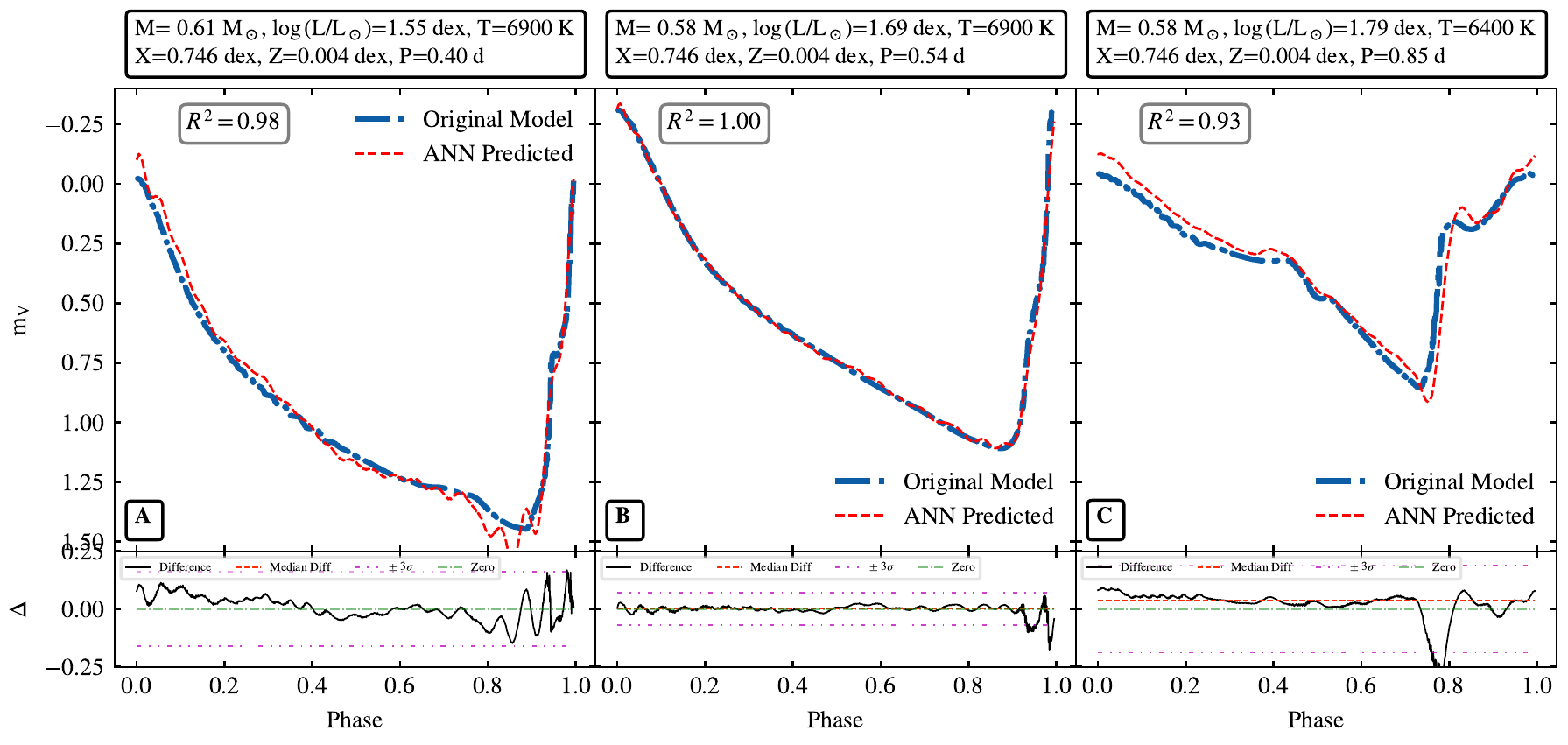}
    \caption{Same as Fig.~\ref{fig:validation_I} but for $V$ band.}
    \label{fig:validation_V}
\end{figure*}

To validate our method, we computed three additional models using the same hydrodynamical code and compared the light curves predicted by the ANN with the ones obtained from these models. The comparison between the ANN generated light curves and the model light curves in $I$ and $V$ bands can be seen in Figs \ref{fig:validation_I} and \ref{fig:validation_V} respectively. The physical parameters for the validation models were selected from a relatively scarce parameter space. The results, shown in Table \ref{tab:validation_stats}, indicate that the light curves predicted by the ANN are consistent with the model light curves. The validation models produced an average mean squared error (MSE) of $2.316 \times 10^{-3}$ in the $I$ band, corresponding to an average $R^{2}$ value of $0.95$, and an MSE of $2.658 \times 10^{-3}$ in the $V$ band, corresponding to an average $R^{2}$ value of $0.97$. This agreement between the ANN generated light curves and the new model light curves demonstrates the validity of our ANN models and confirms that they can be used to generate light curves in both $V$ and $I$ bands.

\begin{table}
    \centering
    \caption{The statistics of the interpolators on validation data.}
    \begin{tabular}{cccccc}
    \hline
         Band & No. of Models & MSE & MAE      & $R^{2}$ \\
     \hline
         I    & 3          & $2.316 \times 10^{-3}$  & $ 3.375 \times 10^{-2}$& $ 0.95123$ \\
         V    & 3          & $2.658 \times 10^{-3}$  & $ 3.5001 \times 10^{-2}$& $ 0.96911$ \\
    \hline
    \end{tabular}
    \label{tab:validation_stats}
\end{table}

\section{Comparison of observed light curves with the generated light curves using interpolators}\label{sec:interpolator_comparison}
Theoretical models are used to complement observational data and are tested on the observed light curves of stars for which the physical parameters are already known. To accurately determine the physical parameters of RR Lyrae stars, a detailed spectroscopic and photometric analysis is often required \citep[see e.g.][]{wang_asteroseismology_2021}. However, in some cases, it is not feasible to obtain spectroscopic measurements of such stars, so data-driven methods are necessary to infer their physical properties. One such method is presented in the study by \cite{bellinger_when_2020}. The authors have predicted the physical parameters (mass, luminosity, effective temperature, and radius) of stars in the LMC and SMC using the OGLE-IV survey, which provides the $I$ and $V$ band light curves of various types of variable stars, including RR Lyrae. The authors used an ANN to derive the parameters by training it with the relationship between light-curve structure (including amplitudes, acuteness, skewness, and coefficients of the Fourier series) and physical parameters of the models. By comparing the ANN generated light curve based on these parameters to the observed light curves, the validity of the derived parameters can be evaluated. This comparison was done by comparing the amplitudes, Fourier parameters, and their distribution with the period of pulsation of the ANN generated light curves to the observed light curves. We stress that the ANN used by \cite{bellinger_when_2020} to estimate the physical parameters based on the light curves and the one we use to predict the light curves based on the physical parameters, were trained on the exact same set of hydrodynamical models.

The ANN model requires a total of six input parameters: $\left[\frac{\text{M}}{\msun}, \,  \log \left(\frac{\text{L}}{\text{L}_{\sun}}\right), \, \teff, \, \text{X}, \, \text{Z}, \, \log (P) \right]$ to predict the light curve. We used the `Lomb-Scargle' algorithm \citep{lomb_least-squares_1976, scargle_studies_1982} to determine the period from the observed magnitudes in un-evenly spaced observations and we calculated the Z values for the \cite{bellinger_when_2020} stars from photometric metallicities ($\feh$) provided by \cite{skowron_ogle-ing_2016}. The steps for transformation from $\feh$ to Z are provided as follows: If the composition of a star is solar scaled, the following relation holds \citep{piersanti_l_method_2007},
\begin{equation}
    \feh = \log (\rm Z/\rm X)_{\star} - \log (\rm Z_{\sun}/\rm X_{\sun}) .
\end{equation}
For Sun, we adopted $\rm X_{\sun}=0.7392, \rm Y_{\sun}=0.2486, \rm Z_{\sun}=0.0122$ from \cite{asplund_solar_2005}. Also, we fixed $ \rm Y =0.245$ for the RRL stars and determined the values of X and Z for each star by cross-matching \citeauthor{bellinger_when_2020} and \citeauthor{skowron_ogle-ing_2016}.

We managed to compile the required input parameters for the $7789$ RRab stars of LMC, and $676$ stars of SMC. With the adopted input parameters we generated the light curves and determined the peak-to-peak amplitude ($A$) and Fourier parameters ($R_{21}, R_{31}, R_{41}, R_{51}, \phi_{21}, \phi_{31}, \phi_{41}, \phi_{51} $) by fitting a Fourier series defined by equation \ref{eq:fourier_sine} with $N=5$.

\begin{figure*}
    \centering
    \includegraphics{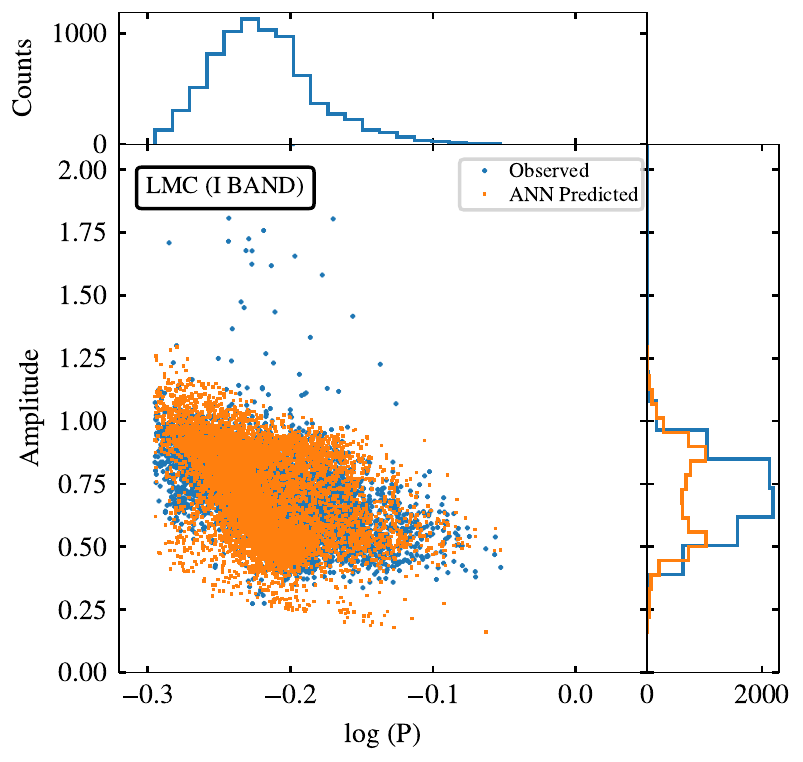}
    \includegraphics{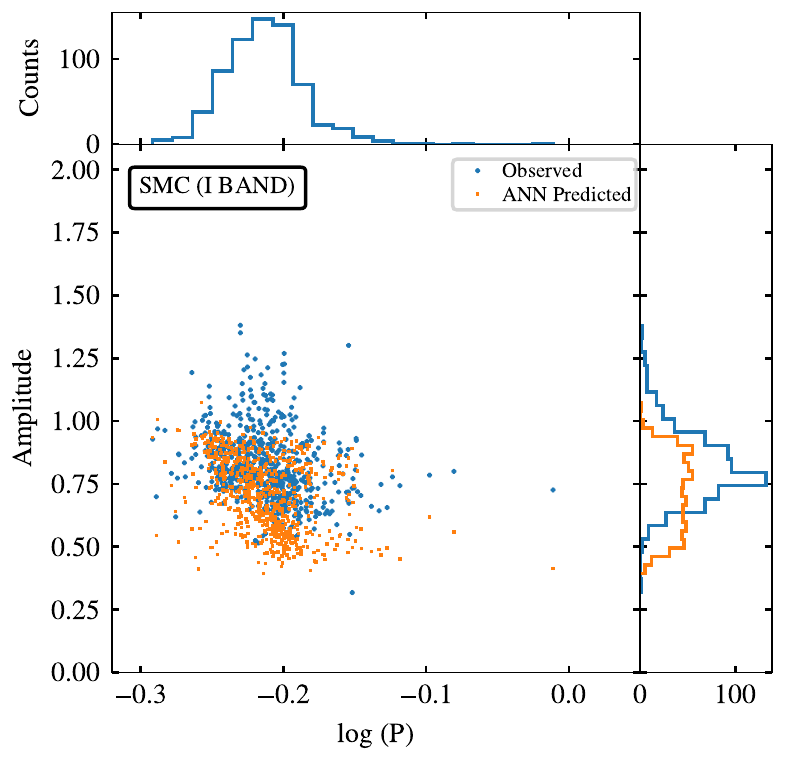}
    \includegraphics{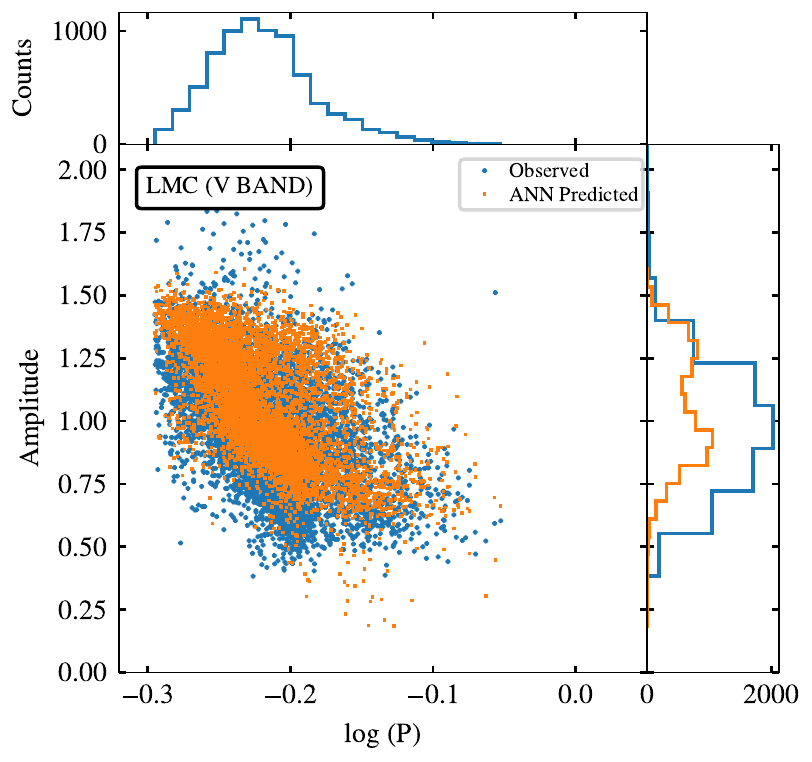}
    \includegraphics{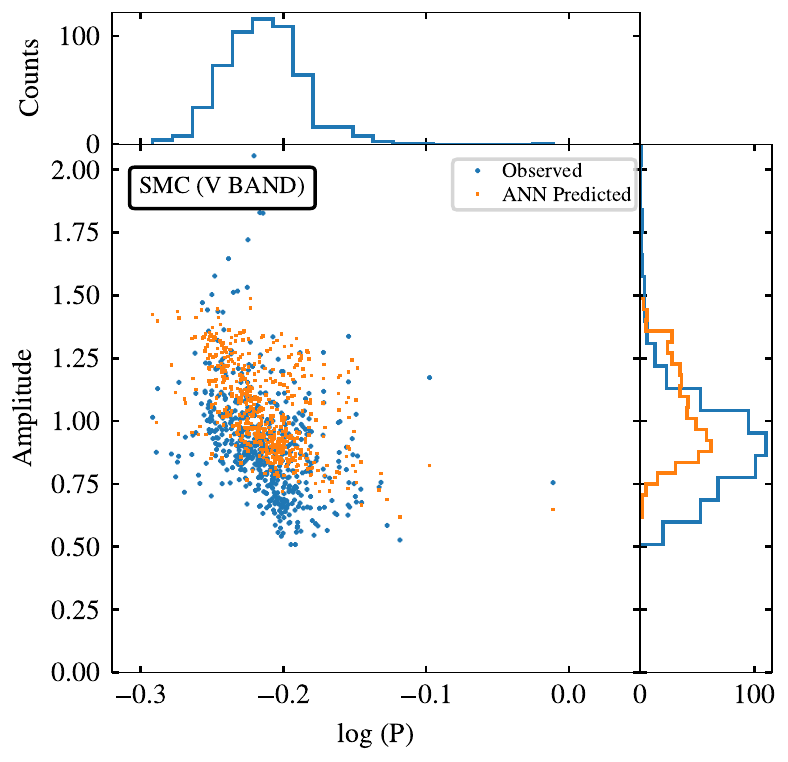}
    \caption{Peak to peak amplitude against $\log(P)$ for Observed and ANN generated light curves of LMC and SMC stars. The extended panel in each plot shows the histograms of periods on the x-axis and amplitudes on the y-axis.}
    \label{fig:period_amplitude}
\end{figure*}

Fig.~\ref{fig:period_amplitude} displays the peak-to-peak amplitudes for the observed and predicted light curves of RR Lyrae in the Magellanic Clouds in the $I$ and $V$ band, respectively. We find that the predicted amplitudes are in good agreement with the observed amplitudes of RR Lyrae variables. For the LMC variables, there seem to be two amplitude sequences for the longer period RRab ($\log \rm P \gtrsim -0.22$) stars. The theoretical models reproduce relatively larger amplitudes for Cepheids and RR Lyrae than the observations in optical bands \citep{bhardwaj_comparative_2017, das_variation_2018}, and the higher amplitude sequence can be attributed to this systematic in the models for specific mass-luminosity levels. The discrepancy is known to be related to the treatment of super adiabatic convection as the considered pulsation models, and \citealt{marconi_new_2015}, assume a single value for the mixing length parameter that is used in the code to close the nonlinear equation system. Nevertheless, the majority of the predicted and the observed amplitudes are in good agreement.

\subsection{Comparison of the Fourier parameters of models with observations}
\begin{figure*}
    \centering
    \includegraphics[scale=0.90]{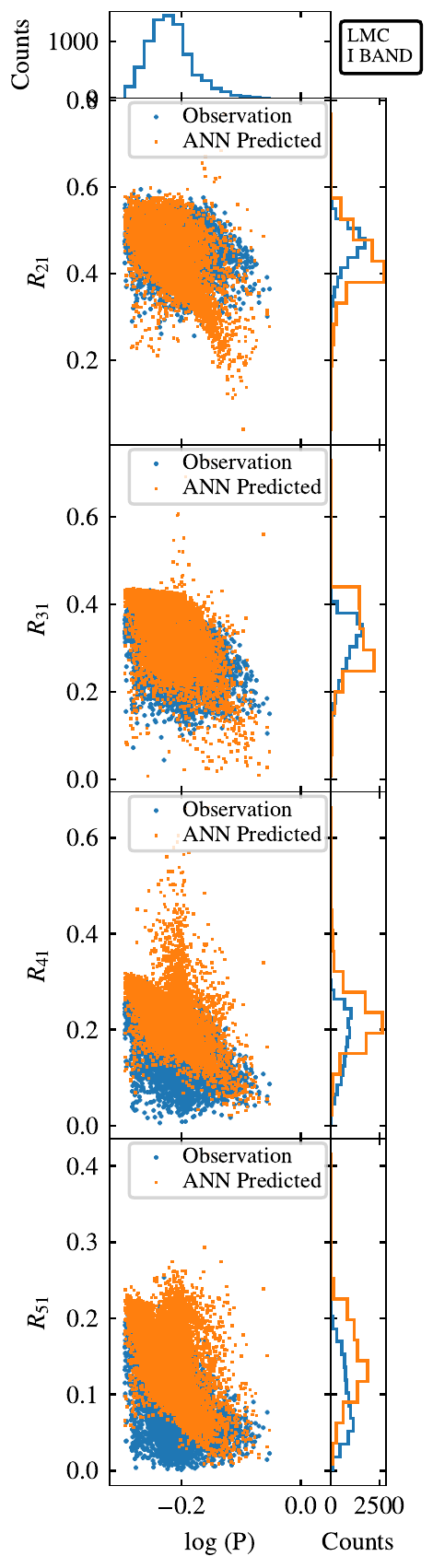}
    \includegraphics[scale=0.90]{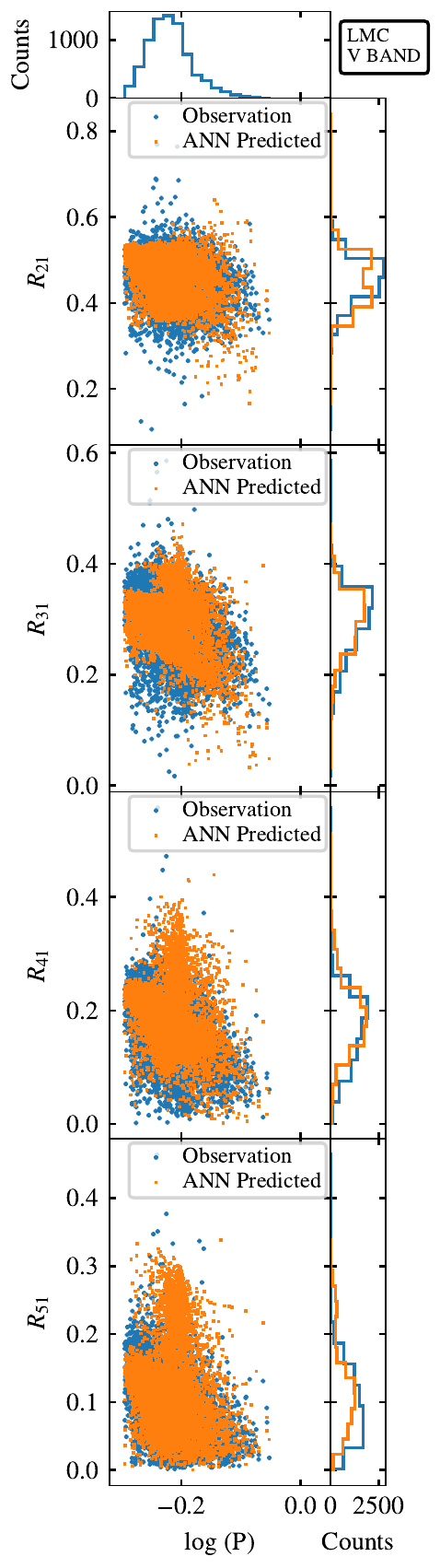}
    \includegraphics[scale=0.90]{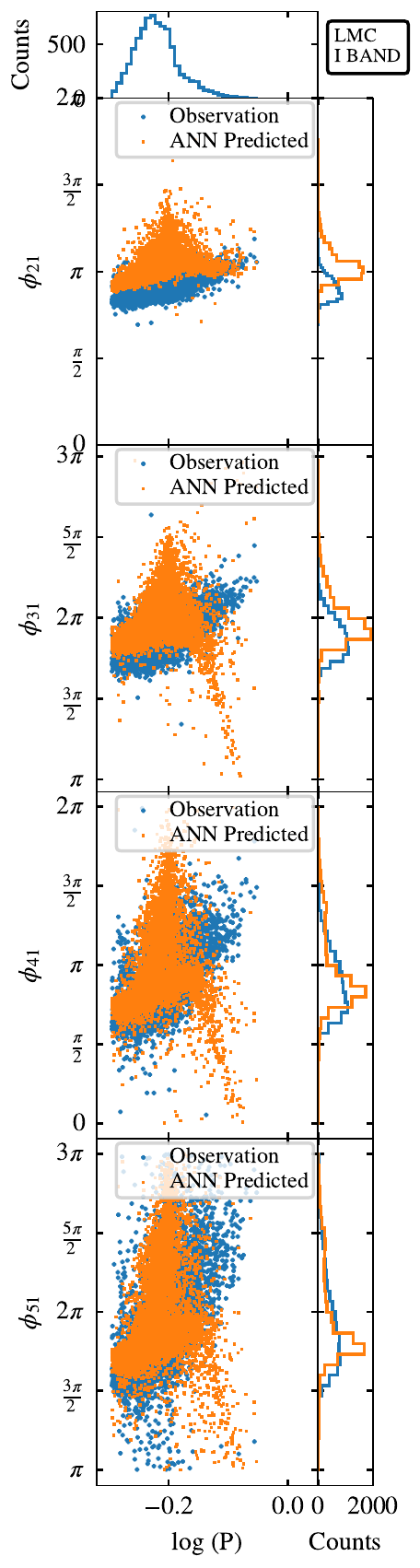}
    \includegraphics[scale=0.90]{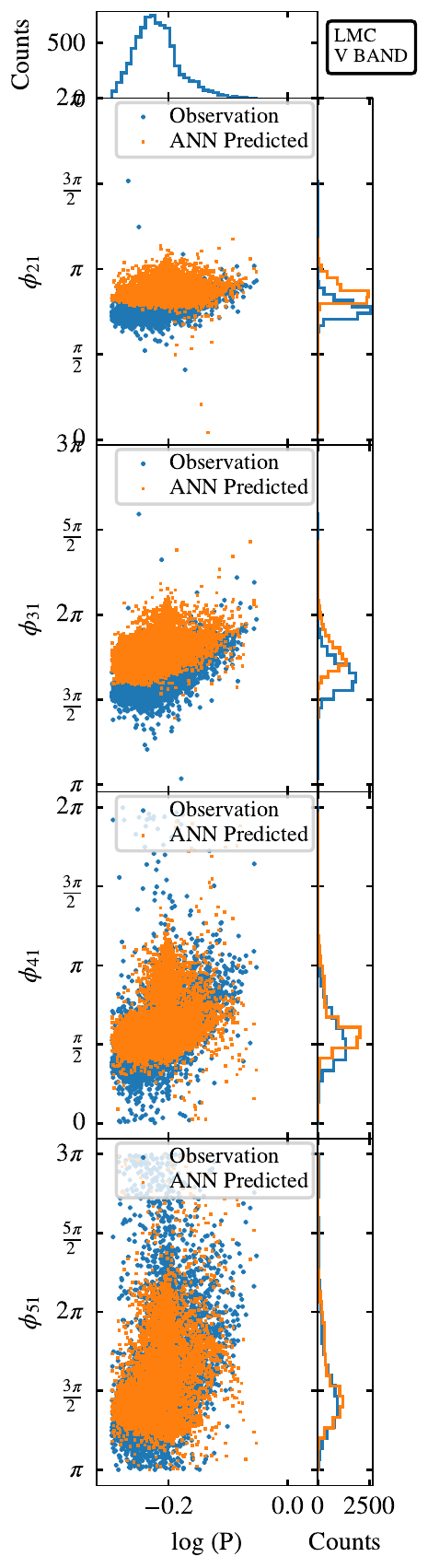}
    \caption{This plot displays the comparison between the Fourier amplitude ratios and phase differences of the light curves predicted by ANN and the actual observations for LMC, in both the $I$ and $V$ bands, as a function of period.}
    \label{fig:FP_compare_LMC}
\end{figure*}

\begin{figure*}
    \centering
    \includegraphics[scale=0.90]{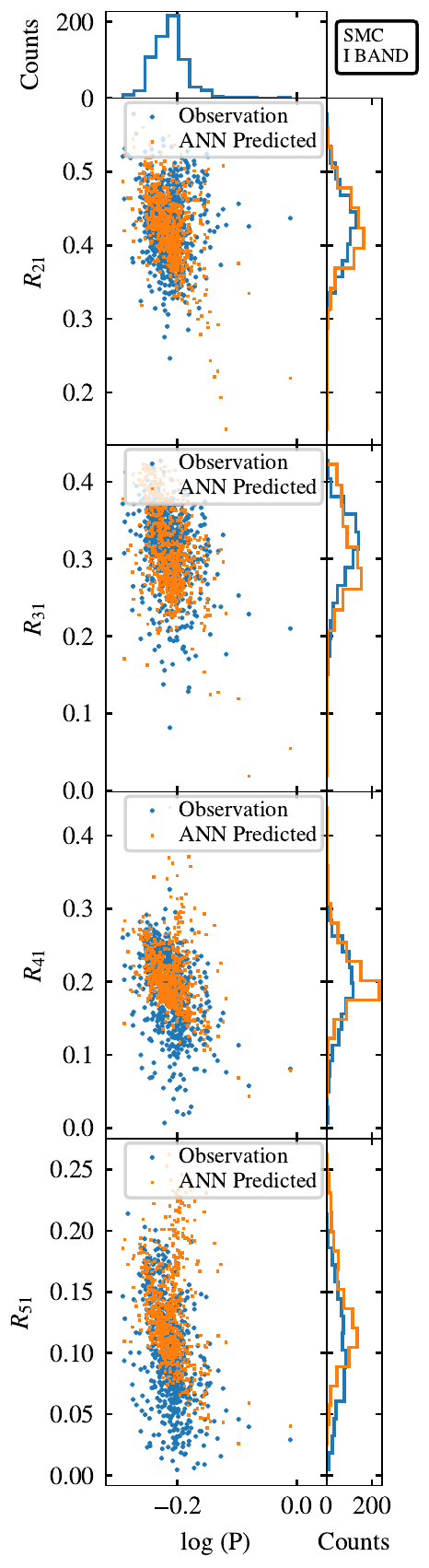}
    \includegraphics[scale=0.90]{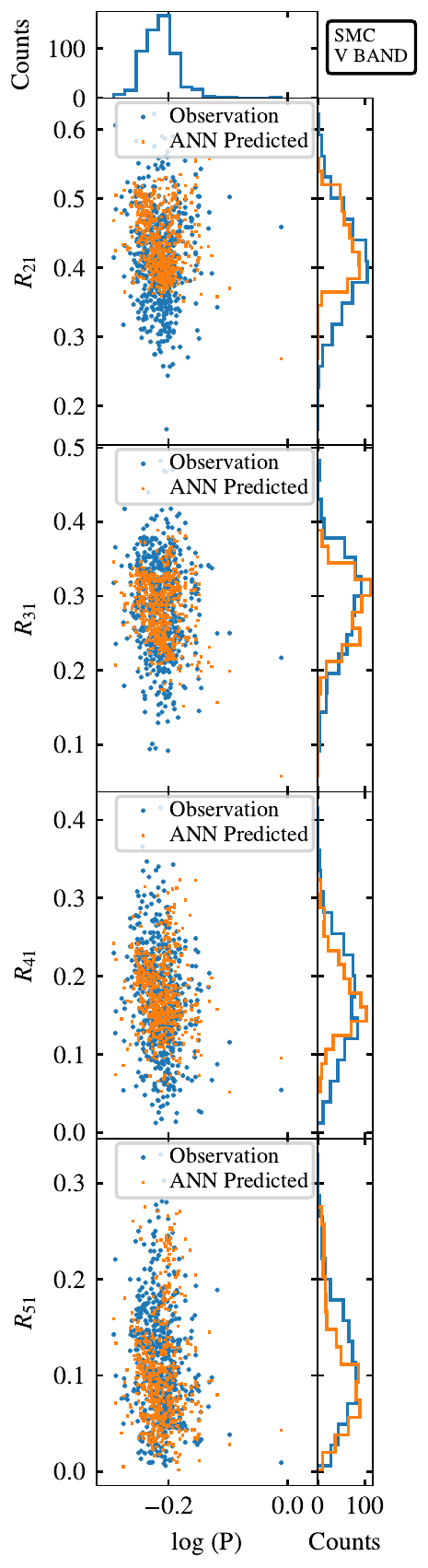}
    \includegraphics[scale=0.90]{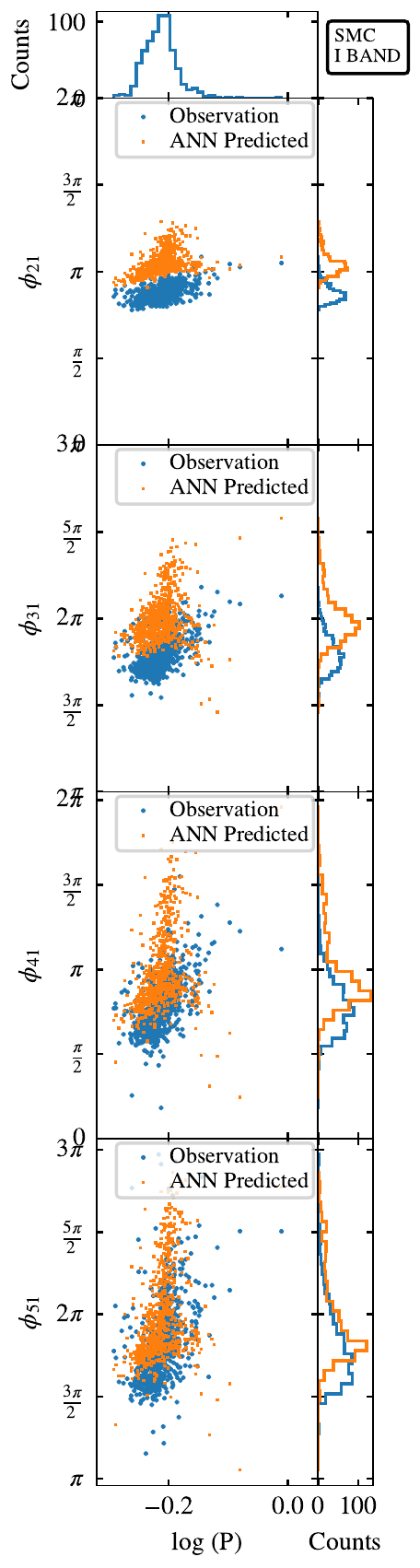}
    \includegraphics[scale=0.90]{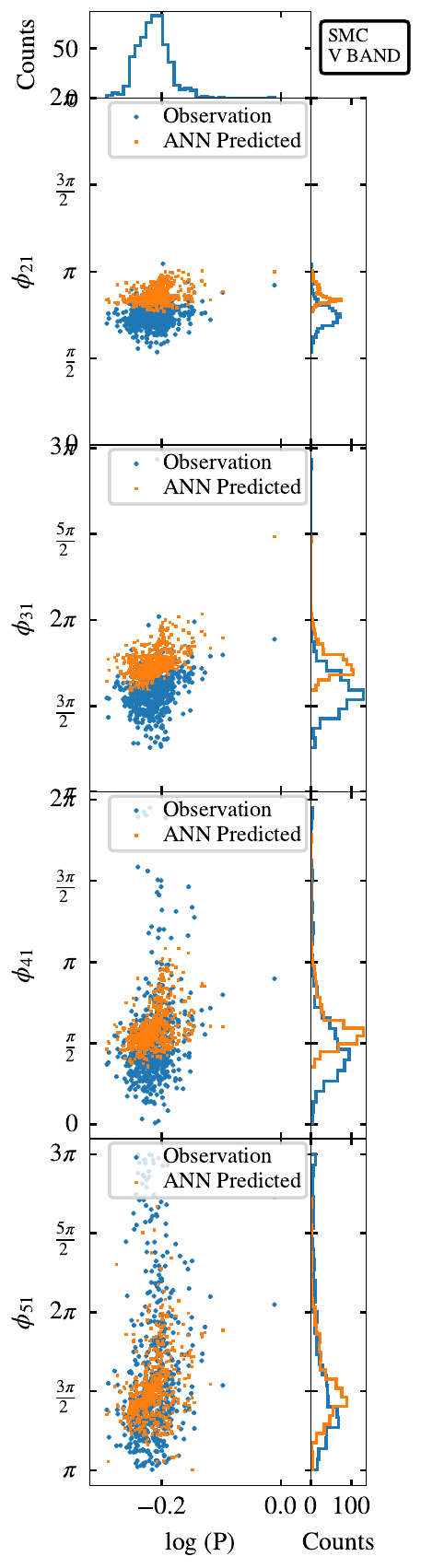}
    \caption{Same as Fig.~\ref{fig:FP_compare_LMC} but for SMC.}
    \label{fig:FP_compare_SMC}
\end{figure*}

Fig.~\ref{fig:FP_compare_LMC} and Fig.~\ref{fig:FP_compare_SMC} display the Fourier parameters of the predicted and observed light curves of LMC and SMC in both $I$ and $V$ bands respectively. For RRab in both the clouds, the Fourier amplitude ratio values from the predicted light curves agree well with the observations, but the predicted phase parameters show a systematic offset and a larger dispersion. We also note that phase parameters exhibit a strong correlation with the metallicity \citep{jurcsik_determination_1996, nemec_metal_2013, mullen_metallicity_2021} and such systematic may also arise from the uncertainties in the input photometric metallicities. 

It should be noted that the predicted physical parameters for these variables from \cite{bellinger_when_2020} are not highly precise, and their accuracy is limited by the lack of a fine grid of models. The derived physical parameters are used to generate the light curve using the trained interpolators. A good match between ANN generated and observed light curves is expected since the same theoretical models were employed to train the ANN used by us and the ANN trained by \cite{bellinger_when_2020}. However, it should be noted that the phase parameters were not included in the training input used to derive the physical parameters in \cite{bellinger_when_2020}. Despite including convection in the hydrodynamical models, it remains difficult to match the observed Fourier phase parameters of RRLs with theoretical models \citep{feuchtinger_nonlinear_1999, paxton_modules_2019}. Comparative studies of the theoretical RR Lyrae models generated by \cite{marconi_new_2015} with the observations show an offset in the value of Fourier phase parameters. The models predict higher Fourier phases than the observations \citep{das_variation_2018}. The same effect gets propagated through the ANN, and the phase parameters derived from the light curves generated using the ANN interpolator are slightly higher than the observed values for both LMC and SMC.


\subsection{Distance modulus to the Magellanic Clouds}\label{sec:distance}
The predicted light curves generated by the ANN model can be employed to estimate the distance modulus of the Magellanic Clouds. We adopted the reddening-independent Wesenheit index \citep{madore_period-luminosity_1982} to determine the distances, which is defined as $W = I - 1.55 \times (V - I)$. The Wesenheit index is a commonly used method for estimating distances \citep{soszynski_ogle_2018}. We calculated the Wesenheit index-based distance moduli ($W_{m} - W_{M}$) to estimate the distances to the Magellanic Clouds.

\begin{figure*}
    \centering
    \includegraphics{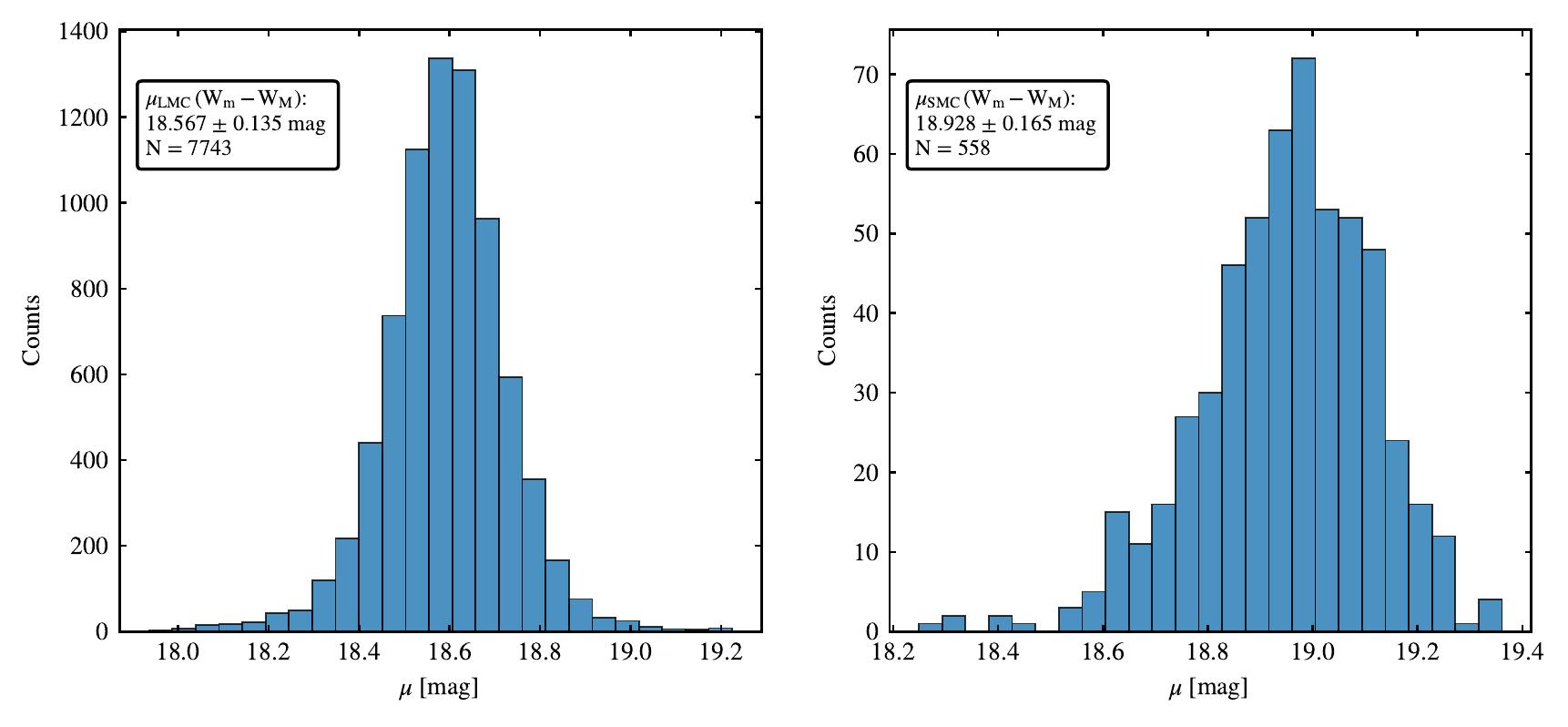}
    \caption{The graph presents a histogram of the distance modulus calculated through the Wesenheit index. The left side shows the distribution of the distance modulus for stars in the Large Magellanic Cloud, while the right side displays the distribution for the Small Magellanic Cloud. In each plot, on the top left corner, a box lists the weighted average distance modulus and the number of stars ($N$) considered after outliers were removed using a 5 $\sigma$ threshold in sigma clipping.}
    \label{fig:distance_LMC_SMC}
\end{figure*}

The distance moduli of individual RRab stars in the Magellanic Clouds are computed based on the Wesenheit index and are depicted in Fig.~\ref{fig:distance_LMC_SMC}. By removing outliers beyond 5-$\sigma$, the average distance modulus of RRab stars in the LMC and SMC are determined to be $ \mu_{\rm LMC} = 18.567 \pm 0.135 $ mag and $ \mu_{\rm SMC} = 18.93 \pm 0.17 $ mag respectively. These estimates are consistent with previously published distances of the Magellanic Clouds based on eclipsing binaries ($ \mu_{\rm LMC} = 18.476 \pm 0.024 $ mag, and $ \mu_{\rm SMC} = 18.95 \pm 0.07$ mag; \citealt{graczyk_araucaria_2014, pietrzynski_distance_2019}). It is important to note that the methodology employed in this study does not presuppose any prior period-magnitude relationship, and the obtained distance estimates are simply a result of accurately predicted light curves of observed stars in the Magellanic Clouds.

\begin{landscape}
    \begin{table}
        \centering
        \caption{The table displays the calculated ANN predicted mean magnitudes, amplitudes, and Fourier parameters, along with the observations. The complete table can be accessed online in a machine-readable format.}        
        \label{tab:compare_SLMC_IV}
        \resizebox{\columnwidth}{!}{\input{table/MERGED_TABLE_1.tex}}
    \end{table}
\end{landscape}

\section{Applications of ANN Interpolators}\label{sec:application}
\subsection{Light Curve Comparison of EZ Cnc}\label{sec:lc_compare_ez_cnc}
As an application to the ANN interpolator, we generated and compared the light curve of an RRab star `EZ Cnc' or `EPIC 212182292' with the observed light curve. It is a non-Blazko RRab variable star which has been observed extensively in both photometric and spectroscopic observing regimes. \cite{wang_asteroseismology_2021} used $55$ high-quality Large Sky Area Multi-Object Fiber Spectroscopic Telescope \citep[LAMOST,][]{luo_first_2015} spectra of medium resolution to determine the atmospheric parameters ($\teff$, $\logg$, and $[\text{M/H}]$). Starting from these parameters, they generated a grid of theoretical models using MESA, applying the time-dependent turbulent convective models. They searched for the optimum parameters for which the modelled light curve matched with the observed light curve from the K2 mission \citep{howell_k2_2014} of the {\it Kepler} spacecraft for which the light curve was processed by EPIC Variability Extraction and Removal for Exoplanet Science Targets Pipeline \citep[EVEREST;][]{luger_everest_2016}. The estimated parameters of the star are given in Table~\ref{tab:ezcnc_params}. The given flux was converted to the {\it Kepler} magnitude (${Kp}$) by formula given by \cite{nemec_fourier_2011},
$$ Kp = m_0 - 2.5 \log (\rm{Flux})$$
where $m_0 = 25.4$ is derived by taking the difference between the instrument magnitude and the mean of $K_p$ \citep{wang_asteroseismology_2021}. We converted the $K_{p}$ to the $V$ band magnitude using the relation given by \cite{nemec_fourier_2011},
$$ V = (1.45 \pm 0.24) Kp - (5.97 \pm 3.20) $$
After extracting the $V$ band light curve, we did extinction correction using the reddening maps of \cite{schlegel_maps_1998, schlafly_measuring_2011} and found E(B-V) $ = 0.027 \pm 0.0006$ and the corresponding $A_V = 3.32 \times 0.0272 \pm 0.0006 = 0.0903 \pm 0.0020$ \footnote{The values are determined from NASA/IPAC INFRARED SCIENCE ARCHIVE (\url{https://irsa.ipac.caltech.edu/applications/DUST/}).} using equation \ref{eq:Av}.

\begin{table}
\centering
\caption{The stellar parameters of EZ Cnc (EPIC 212182292) adopted from \citet{wang_asteroseismology_2021}}.
\label{tab:ezcnc_params}
\resizebox{\linewidth}{!}{%
\begin{tabular}{>{\hspace{0pt}}m{0.458\linewidth}>{\hspace{0pt}}m{0.485\linewidth}} 
\hline
Parameter & Value \\ 
\hline
Mass (M) & $0.48 \pm 0.03 \ \msun$ \\
Luminosity (L) & $42 \pm 2 \ \lsun$ \\
Effective temperature ($\teff$) & $6846 \pm 50$ K \\
X & $0.741 \pm 0.004$ dex \\
Z & $0.006 \pm 0.002$ dex \\
Period (P)$\rm^a$ & $0.545740 \pm 0.000007$ days \\ 
\hline
\multicolumn{2}{>{\hspace{0pt}}m{0.943\linewidth}}{\begin{tabular}[c]{@{}l@{}}$\rm^{a}$The period is determined using the Kepler light-curve with\\~ ~ ~ ~ \texttt{PERIOD-04} \citep{lenz_period04_2005}.\end{tabular}}
\end{tabular}}
\end{table}

\begin{figure}
    \centering
    \includegraphics{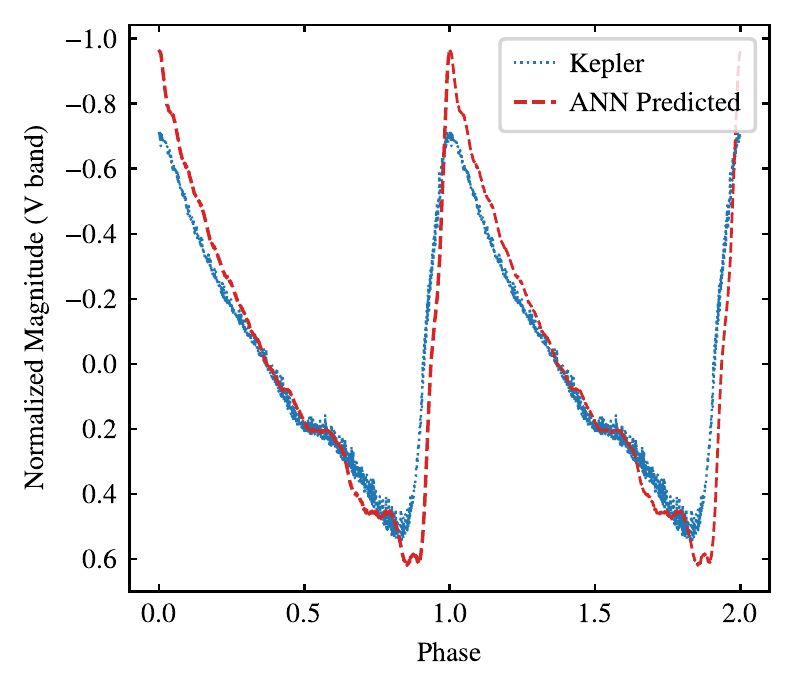}
    \caption{This plot represents the observed light curve from the K2 survey (converted to $V$ band) along with the ANN generated light curve using the trained $V$ band interpolator.}
    \label{fig:ez_cnc}
\end{figure}

We generated a light curve using the $V$ band interpolator by providing the parameters given in Table~\ref{tab:ezcnc_params} as input. The error bars for the predicted magnitude are derived from the given uncertainties in the physical parameters. The resulting ANN generated light curve is plotted against the observed $V$ band light curve from the {\it Kepler} telescope in Fig.~\ref{fig:ez_cnc}. For the purpose of comparison, both light curves are normalised to mean magnitudes. We also did the quantitative analysis by determining and comparing the Fourier amplitude and phase parameter ratio for the \textit{Kepler} $V$ band, and ANN predicted $V$ band light curve. The derived parameters including amplitudes and Fourier parameter ratios are given in Table~\ref{tab:ez_cnc_fp}. The peak-to-peak amplitude (A) predicted by the ANN $V$ band interpolator is higher than the observed $V$ band light curve. This is due to a rather low mixing length parameter adopted in the considered grid of models. For a successful model fitting applications to RR Lyrae stars \citep[see e.g. ][]{marconi_pulsational_2005, marconi_modeling_2007}, an increased mixing length value is required to match the light curves of fundamental pulsators. The Fourier amplitude and phase ratio of the K2 and ANN generated light curves match each other within the $1\ \sigma$ errors. 

A direct application of this analysis is to estimate the distance to the star. Since we have calculated the mean absolute magnitude from the ANN generated light curve and the apparent magnitude from the \textit{Kepler} telescope, the distance modulus is calculated as $ \mu_{\rm EZ\ Cnc} = 11.988(2) - 0.703(2) = 11.284(3)$ mag, and hence the distance $d$ (in pc) calculated using the $\mu = 5 \log(d) - 5 $ gives $d = 1806 \pm 2 $ pc. The calibrated {\it Gaia} DR2 distance for this star is $1840^{+192}_{-161}$ pc \citep{bailer-jones_estimating_2018}, which is recently updated by {\it Gaia} EDR3 to $1775^{+70}_{-70}$ pc \citep{bailer-jones_estimating_2021}. The estimated distance to the EZ Cnc remarkably matches with the published distance estimations in the literature.

\begin{table}
    \centering
    \caption{Comparison of the Fourier parameters of EZ Cnc star derived using K2 and ANN predicted light curve.}
    \resizebox{\columnwidth}{!}{%
    \begin{tabular}{lcc}
    \hline
          & K2 light curve  &  ANN Predicted  \\
          & ($Kp$ band converted to $V$ band) &  ($V$ band) \\
    \hline
         A (mag) & $1.263 \pm 0.148$ & $1.584 \pm 0.107$ \\
         mean mag (mag) &  $11.988 \pm 0.002$ & $0.703 \pm 0.002$ \\
         $R_{21}$ & $0.516 \pm 0.002$ & $0.542 \pm 0.005$ \\
         $R_{31}$ & $0.327 \pm 0.002$ & $0.325 \pm 0.005$\\
         $R_{41}$ & $0.153 \pm 0.002$ &  $0.223 \pm 0.005$\\
         $R_{51}$ & $0.104 \pm 0.002$ & $0.178 \pm 0.005$ \\
         $\phi_{21}$ & $2.761 \pm 0.005$ & $2.783 \pm 0.013$ \\
         $\phi_{31}$ & $5.719 \pm 0.007$ & $5.718 \pm 0.020$ \\
         $\phi_{41}$ & $2.214 \pm 0.013$ & $2.170 \pm 0.027$\\
         $\phi_{51}$ & $5.024 \pm 0.018$ & $5.119 \pm 0.034$\\
    \hline
    \end{tabular}
    }
    \label{tab:ez_cnc_fp}
\end{table}

\subsection{Generating a grid of models using the trained ANN interpolators}\label{sec:new_grid}
To get the precise physical parameters of pulsating variable stars, a grid of model light curves is compared with the observed light curve. However, a pre-computed grid of models is usually very coarse and non-uniform in the parameter space. This is due to the high computation cost and time consuming process of solving time-dependent hydrodynamical equations of the stellar atmosphere. Moreover, to constrain the parameters like mass, surface gravity and metallicity, we need to rely on spectroscopic data which is usually not available with photometric data. Hence, a fine grid of models is required to pin down the physical and atmospheric parameters of the star.

We generated a fine grid of light curve templates in both $I$ and $V$ bands using the trained interpolators. The choice of input parameters for the new grid is limited by the parameter space of the original grid of models. We choose a finer and more uniform grid than the original models. We generated the grid for three helium abundance ratios Y=$0.245, 0.25,$ and $0.265$ and $4$ different Z values ranging from metal-poor to  metal-rich stars. For any combination of Y and Z, the hydrogen abundance ratio (X) can be calculated using X = 1- Y- Z. Mass (M) varies from $0.52$ to $0.79\, \msun$, with a constant step size of $ 0.03\, \msun$. The luminosity parameter $\log ({\rm L/ \rm L_\odot })$ varies from $1.54$ to $2.02$ dex, with a step size of $0.04$ dex, the effective temperature ($\teff$) ranges from $5300$ to $7000$ K with a step size of $100$ K. The period of an RR Lyrae star is closely related to its temperature, luminosity, and mass \cite{van_albada_1971}. The van Albada-Baker (vAB) relation describes this relationship. We have used a modern version of the vAB relation, which includes the effect of metallicity on the pulsation period, from \cite{marconi_new_2015}. We used the relation for the fundamental mode RRLs.
\begin{equation}
    \begin{aligned}
            \log P =& -(0.58 \pm 0.02) \log\left({\frac{\rm M}{\msun}}\right) + (0.860 \pm 0.003) \log\left({\frac{\rm L}{\lsun}}\right) \\
            & - (3.40 \pm 0.03) \log(\teff) + (0.013 \pm 0.002) \log(Z) \\
                     & + (11.347 \pm 0.006).
    \end{aligned}
\end{equation}

\begin{figure*}
    \centering
    \includegraphics{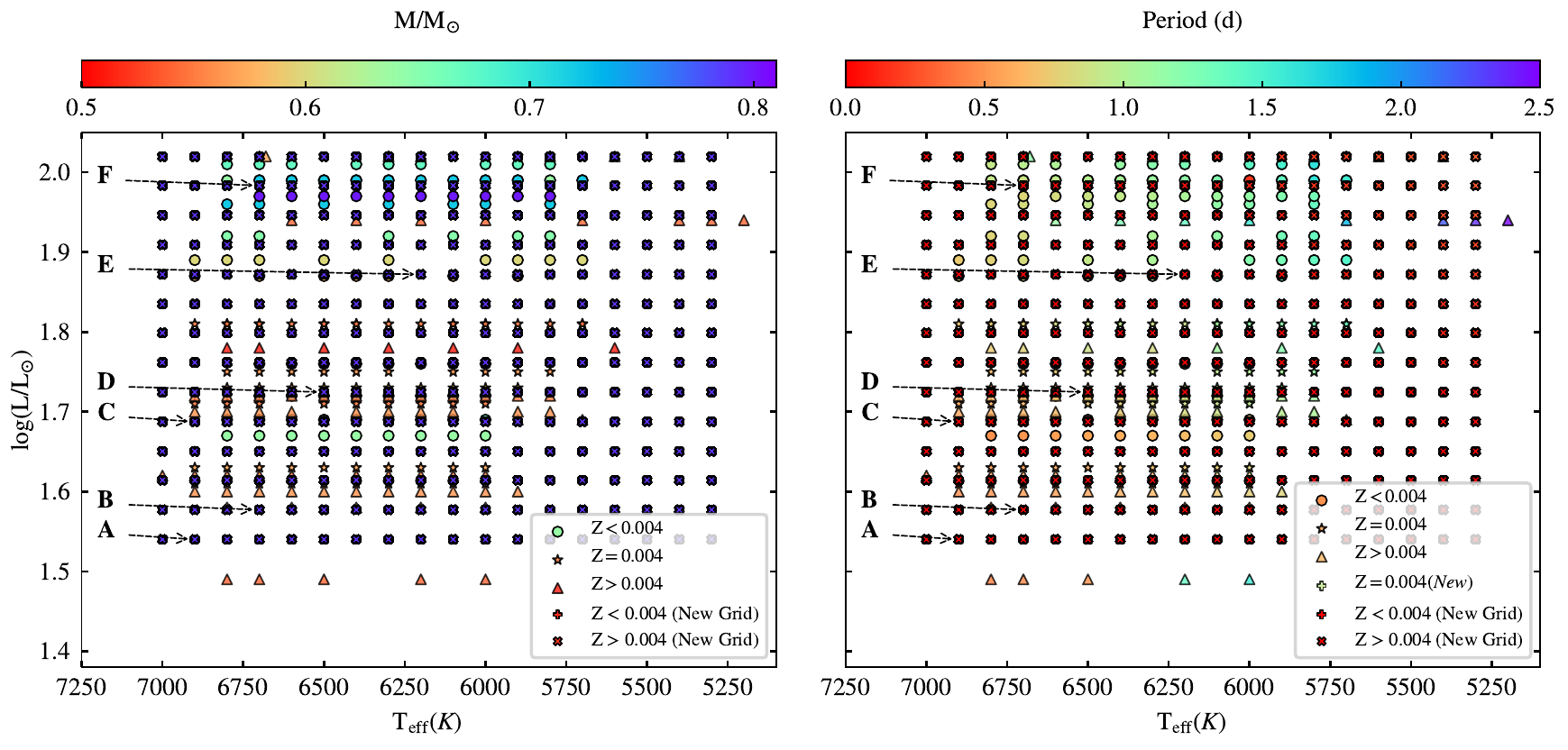}
    \caption{The parameter space of the new grid along with \citet{marconi_new_2015}'s original grid. A more detailed grid can be found in Fig.~\ref{fig:input_distribution_appendix}. The ANN generated light curves of the labelled models are shown in Fig.~\ref{fig:six_stars}.}
    \label{fig:new_grid}
\end{figure*}

\begin{figure*}
    \centering
    \includegraphics{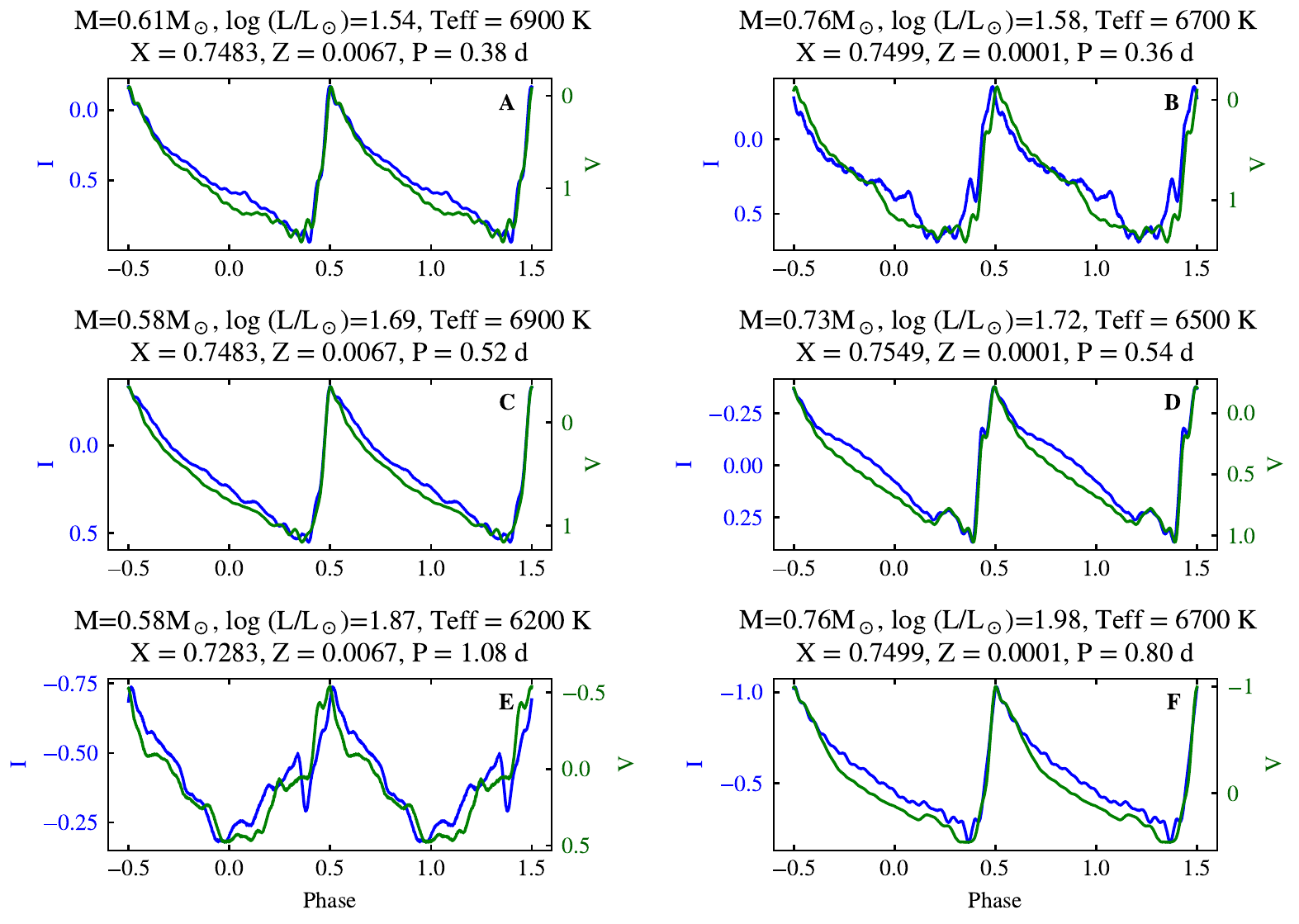}
    \caption{The ANN generated $I$ and $V$ band light curves for six models labelled in Fig.~\ref{fig:new_grid}.}
    \label{fig:six_stars}
\end{figure*}

We end up with a grid of $37,800$ individual parameter combinations for which the template light curves are generated in the $I$ and $V$ band using the interpolators. Fig.~\ref{fig:new_grid} represents the distribution of \cite{marconi_new_2015} parameter space with the new grid parameters that we have computed (see Table~\ref{tab:new_grid} for the parameter ranges). A complete distribution of all parameters is shown in Fig.~\ref{fig:input_distribution_appendix}. The light curve templates of six random models of the new grid are shown in Fig.~\ref{fig:six_stars} in both $I$ and $V$ bands. We observe that the predicted light curves exhibit the same structure and features as an RRab light curve. However, for certain combinations of the input parameters, the predicted light curve may not resemble an RRab light curve. The reason for this can be traced back to the scarcity of models in this region of the training dataset or the lack of stable RRab stars with these combinations of physical parameters.

\begin{table}
    \centering
    \caption{The parameter space of the new grid generated using the trained ANN Interpolators.}
    \label{tab:new_grid}
    \begin{tabular}{ccc}
    \hline
         Parameter   & Range & Step  \\
         \hline
         M  &  0.52 - 0.79 $\msun$     &     0.03 $\msun$   \\
         $\log ({\rm L/ \rm L_\odot })$& 1.54 - 2.02 dex & 0.04 dex\\
         $\teff$ & 5300 - 7000 K & 100 K \\
         Y & \multicolumn{2}{l}{{[}0.245, 0.25, 0.265]}  \\
         Z & \multicolumn{2}{l}{{[}0.00011, 0.00668, 0.01324, 0.01980]}  \\ 
        \hline
    \end{tabular}
\end{table}

\section{Summary}\label{sec:summary}
We present a new technique to generate the light curve of RRab models in different photometric bands using ANN. We built and trained an artificial neural network for interpolating the light curve within a pre-computed grid of models. The ANN has been trained with the physical parameters-light curve grid. We used the models generated by \cite{marconi_new_2015} and used in \cite{das_variation_2018}, which were computed by solving the hydrodynamical conservation equations simultaneously with a nonlocal, time-dependent treatment of convective transport \citep{stellingwerf_convection_1982, bono_pulsation_1994, bono_classical_2000, marconi_2009}. For the validation of the trained interpolators, light curves for a few new models were generated and then compared with the ANN predicted light curves.

The architecture of the neural network is tuned using the $I$ band light curves. A random search for the hyperparameters is performed within a grid of {\it inferred} hyperparameter combinations. The architecture and trained weights of the best-performing network are then adopted for making the final interpolators in $I$ and $V$ bands.  

As an application of the trained interpolators, we generated and compared the light curves of the RRab stars in the Magellanic clouds (LMC/SMC). The physical parameters [M/$\msun, \log (\rm L/\lsun), \teff$] of RRab stars in LMC and SMC are adopted from \cite{bellinger_when_2020}. Z is calculated from the metallicity estimates provided by \citet{skowron_ogle-ing_2016}, and X $= 1 - $Y $-$ Z; is calculated using a fixed primordial helium abundance (Y$ = 0.245$). Lastly, the period is determined from the observed light curve using the `Lomb-scargle' method. The interpolators are then used to predict the light curve from the given physical parameters. Both observed and predicted light curves are then fitted with a Fourier sine series (see eq. \ref{eq:fourier_sine}) with N=5. The comparison of ANN predicted amplitudes, Fourier amplitude ratios and Fourier phase differences with the observations is shown in Fig.~\ref{fig:period_amplitude}, ~\ref{fig:FP_compare_LMC}, \& ~\ref{fig:FP_compare_SMC}. We observe that ANN predicted amplitudes are a bit larger than the observed amplitudes and the reason for this can be traced back to the low mixing length that was used to compute the original models. The Fourier amplitude ratios ($R_{k1}$, as defined in eq.~\ref{eq:Rk1}) of the light curves predicted by the ANN are in great agreement with observations, except for a few exceptions where $R_{41}$ and $R_{51}$ exhibit an additional feature in the ANN predicted models at around $\log(P) \approx -0.22$. The Fourier phase differences ($\phi_{k1}$, defined in equation~\ref{eq:phi_k1}) are consistently shifted, particularly $\phi_{21}$ and $\phi_{31}$. The cause of this discrepancy is not understood and it should be noted that even the state-of-the-art hydrodynamical code, \texttt{MESA-rsp}, is unable to accurately reproduce the Fourier phase differences \citep{paxton_modules_2019}. We also determine the distances to the LMC and SMC based on the reddening independent Wesenheit index. The distances found ($ \mu_{\rm LMC} = 18.567 \pm 0.135 $ mag, and $ \mu_{\rm SMC} = 18.93 \pm 0.17$ mag) are in excellent agreement with the published distances based on eclipsing binaries ($ \mu_{\rm LMC} = 18.476 \pm 0.024 $ mag, and $ \mu_{\rm SMC} = 18.95 \pm 0.07$ mag; \citealt{graczyk_araucaria_2014, pietrzynski_distance_2019}).

To showcase the utility of the interpolators, we generated and compared the light curve of the RRab star EZ Cnc. The physical parameters of this star were determined by \cite{wang_asteroseismology_2021} using medium-resolution spectroscopic observations from LAMOST and time-series photometric data from the {\it Kepler} mission. We transformed the {\it Kepler} light curve into the $V$-band light curve and then compared it with the light curve predicted by the ANN both qualitatively and quantitatively. The reported distance to this star, $ \mu_{\rm EZ\ Cnc} = 11.284(3)$, or $d = 1806 \pm 2 $ pc, is in excellent agreement with the recently updated parallax measurement from {\it Gaia} EDR3 of $1775^{+70}_{-70}$ pc \citep{bailer-jones_estimating_2021}.

The generation of a grid of model light curves using traditional methods can be computationally expensive and time-consuming, but by using the trained ANN interpolators, it is possible to generate a more dense grid of model light curves much more efficiently. The trained interpolators can generate a light curve given the input parameters, and the process is fast, taking only a few milliseconds for each light curve. Additionally, the size of the trained interpolator file is much smaller, making it easy to store and access. To complement existing theoretical model grids in the literature, a smooth grid of model light curves was generated using the trained interpolator. The grid of templates can be used in techniques such as template fitting to estimate the parameters of observed light curves. We generated over 30,000 model light curves in both the $I$ and $V$ bands, resulting in approximately 2 GB of data. However, if one has access to the trained interpolator file, which is much smaller in size (around 3.7 MB for each interpolator file in our case), it is also possible to generate a light curve by inputting the parameters. Generating each light curve takes only a few milliseconds (approximately 55 ms) for both $I$ and $V$ bands.

It is worth noting that our approach is dependent on the models used, and any errors or uncertainties in the models will be reflected in our results. However, this analysis will provide valuable insights into the stellar population model and has the potential to improve our understanding of these stars. The results can be improved by expanding the number of models or by using a more comprehensive grid of models. Additionally, the trained ANN models can be retrained on new or additional models to enhance the accuracy of the predicted light curves. In this way, our approach can be continuously refined and improved as more data and models become available.

\section*{Software}
We utilized various Python libraries in our study including \texttt{Numpy} \citep[][]{harris_array_2020}, \texttt{Pandas} \citep[][]{mckinney_data_2010, team_pandas-devpandas_2020}, \texttt{Astropy}\footnote{\url{http://www.astropy.org}}\citep{astropy_collaboration_astropy_2013, astropy_collaboration_astropy_2018}, \texttt{TensorFlow} \citep{martin_abadi_tensorflow_2015}, \texttt{Matplotlib} \citep[]{hunter_matplotlib_2007} and \texttt{Seaborn} \citep{waskom_seaborn_2021}. \texttt{Numpy} and \texttt{Pandas} were used for data manipulation, while \texttt{Astropy}, a community-developed package for Astronomy, was also utilized. \texttt{TensorFlow} was employed to implement the Artificial Neural Network, while \texttt{Matplotlib} and \texttt{Seaborn} were used for creating visual plots.

\section*{Acknowledgements}
NK acknowledges the financial assistance from the Council of Scientific and Industrial Research (CSIR), New Delhi, as the Senior Research Fellowship (SRF) file no. 09/45(1651)/2019-EMR-I. AB acknowledges funding from the European Union’s Horizon 2020 research and innovation programme under the Marie Skłodowska-Curie grant agreement No. 886298. SD acknowledges the KKP-137523 `SeismoLab' Élvonal grant of the Hungarian Research, Development and Innovation Office (NKFIH). HPS acknowledges a grant from the Council of Scientific and Industrial Research (CSIR) India, file no. 03(1428)/18-EMR-II. 

\section*{Data Availability}
Interested readers can use the trained interpolator to generate the predicted light curves of RRab stars using their input physical parameters. The generated grid of light curves is available on request and also through a web interface on \url{https://ann-interpolator.web.app/}.

\bibliographystyle{mnras}
\bibliography{paper} 


\appendix

\section{Input Distribution}
\begin{figure*}
    \centering
    \includegraphics[width=\textwidth]{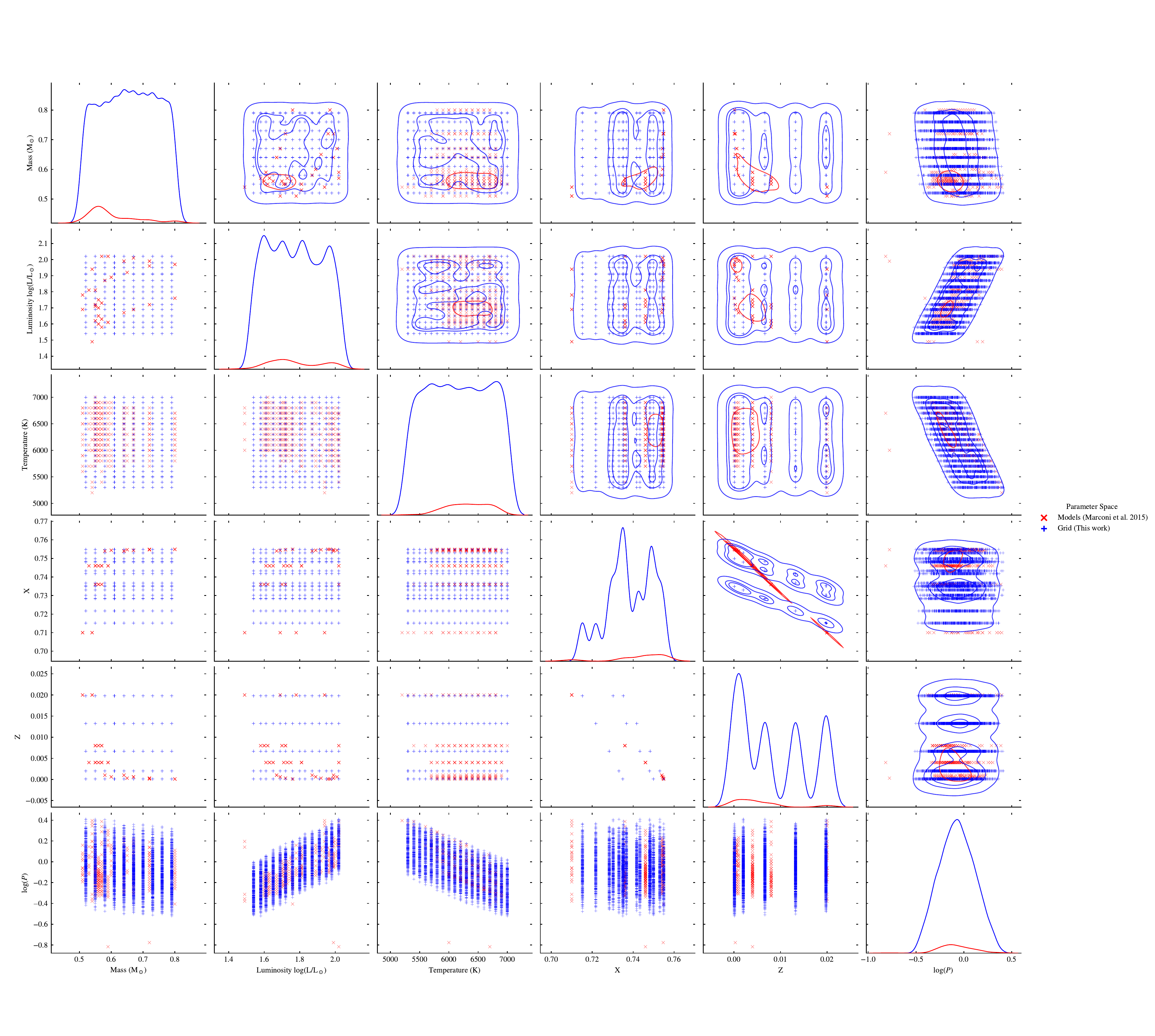}
    \caption{A pair plot with KDE (Kernel Density Estimation) plots in the upper triangle shows the relationship between six variables, namely mass, luminosity, effective temperature, X, Z, and log(P). The contours represent the parameter space in 2D for the given combination of six parameters. The plot is divided into subplots, with each subplot representing a pair-wise scatter plot of the six variables. The diagonal of the subplots shows the distribution of each variable using KDE plots. The red cross (`\textcolor{red}{$\times$}') marker is assigned for the original models of \protect\cite{marconi_new_2015} and the blue plus (`\protect\textcolor{blue}{+}') marker is assigned for the models generated in this work.}
    \label{fig:input_distribution_appendix}
\end{figure*}

A detailed input distribution of the input parameters is shown in Fig.~\ref{fig:input_distribution_appendix}. We have also shown the detailed distribution of parameters of the new grid for which the template light curves are generated. 

\bsp	
\label{lastpage}
\end{document}

%% file: nn_customization.tex
\usepackage{pgfplots}
\pgfplotsset{compat=1.15}

\tikzset{mystyle/.style={red, draw=blue, fill=yellow!20}}

\newcommand{\nodetextclear}[2]{}

\newcommand{\nodetexts}[2]{$s$}
\newcommand{\nodetextsigma}[2]{$\Sigma$}
\newcommand{\nodetextstep}[2]{
    \tikzstyle{input} = [coordinate]
    \begin{tikzpicture}[auto, >=latex']
        \draw[line width=0.05mm](0,0)  +(0,-6pt) -- +(0,6pt); 
        \draw[line width=0.25mm](0,0)  +(0pt,3pt) -- +(6pt,3pt); 
        \draw[line width=0.25mm](0,0)  +(0,-3pt) -- +(0,3pt); 
        \draw[line width=0.25mm](0,0)  +(-6pt,-3pt) -- +(0pt,-3pt); 
    \end{tikzpicture}
}
\newcommand{\nodetextsigmoid}[2]{
    \begin{tikzpicture}[auto, >=latex']
        \begin{axis}[hide axis, scale only axis, height=2ex]
            \addplot[color=black]{1/(1+exp(-x))-2};
        \end{axis}
    \end{tikzpicture}
}

%% file: table/parameter_table.tex
\begin{tabular}{cccccc}
    \hline
    Z & X & $\frac{\rm M}{\rm M_{\odot}}$ & log $\frac{\rm L}{\rm L_{\odot}}$ & $\teff(K)$ & \makecell[t]{No. of\\ RRab stars} \\
    \hline 
    0.02 & 0.71 & 0.51 & 1.69 & 5700 - 6800 & 6 \\ 
    ~ & ~ & ~ & 1.78 & 5600 - 6800 & 7 \\ 
    ~ & ~ & 0.54 & 1.49 & 6000 - 6800 & 5 \\ 
    ~ & ~ & ~ & 1.94 & 5200 - 6600 & 8 \\ \hline
    0.008 & 0.736 & 0.55 & 1.62 & 6000 - 7000 & 10 \\ 
    ~ & ~ & ~ & 1.72 & 5800 - 6900 & 12 \\ 
    ~ & ~ & 0.56 & 1.6 & 5900 - 6900 & 11 \\ 
    ~ & ~ & ~ & 1.7 & 5800 - 6900 & 10 \\ 
    ~ & ~ & 0.57 & 1.58 & 6000 - 6800 & 5 \\ 
    ~ & ~ & ~ & 2.02 & 5400 - 6680 & 6 \\  \hline
    0.004 & 0.746 & 0.53 & 1.81 & 5700 - 6800 & 7 \\ 
    ~ & ~ & 0.55 & 1.71 & 6000 - 6900 & 9 \\
    ~ & ~ & ~ & 1.81 & 5700 - 6900 & 13 \\  
    ~ & ~ & 0.56 & 1.65 & 6000 - 6900 & 10 \\  
    ~ & ~ & ~ & 1.75 & 5800 - 6800 & 10 \\  
    ~ & ~ & 0.57 & 1.63 & 6000 - 6900 & 10 \\  
    ~ & ~ & ~ & 1.73 & 5900 - 6800 & 10 \\  
    ~ & ~ & 0.59 & 1.61 & 6000 - 6900 & 10 \\  
    ~ & ~ & ~ & 2.02 & 5700 - 6700 & 7 \\  \hline
    0.001 & 0.754 & 0.58 & 1.87 & 5900 - 6900 & 7 \\  
    ~ & ~ & 0.64 & 1.67 & 6000 - 6800 & 9 \\  
    ~ & ~ & ~ & 1.99 & 5700 - 6800 & 10 \\  \hline
    0.0006 & 0.7544 & 0.6 & 1.89 & 5700 - 6900 & 9 \\  
    ~ & ~ & 0.67 & 1.69 & 6000 - 6800 & 9 \\  
    ~ & ~ & ~ & 2.01 & 5800 - 6800 & 9 \\  \hline
    0.0003 & 0.7547 & 0.65 & 1.92 & 5800 - 6800 & 6 \\  
    ~ & ~ & 0.72 & 1.72 & 6000 - 6700 & 8 \\  
    ~ & ~ & ~ & 1.99 & 5700 - 6700 & 10 \\  \hline
    0.0001 & 0.7549 & 0.72 & 1.96 & 5800 - 6800 & 7 \\  
    ~ & ~ & 0.8 & 1.76 & 6000 - 6700 & 8 \\  
    ~ & ~ & ~ & 1.97 & 5800 - 6700 & 10 \\  
    \hline
\end{tabular}

%% file: table/hparams_tune.tex
\begin{tabular}{cccccccc}
\hline
Architecture No. & No. of Hidden Layers & Hlayer 1 & Hlayer 2 & Hlayer 3 & Activation & {Learning Rate($\eta$)} & {Min MSE} \\ 
\hline
1 & 3 & 64 & 128 & 128 & relu & 1.7941$\times 10^{-3}$ & 8.7585$\times 10^{-5}$ \\ 
2 & 3 & 32 & 64 & 128 & relu & 2.5664$\times 10^{-3}$ & 1.3403$\times 10^{-4}$ \\ 
3 & 3 & 64 & 32 & 128 & relu & 3.3848$\times 10^{-3}$ & 1.8869$\times 10^{-4}$ \\ 
4 & 3 & 64 & 128 & 128 & relu & 8.2402$\times 10^{-4}$ & 2.1454$\times 10^{-4}$ \\ 
5 & 3 & 128 & 64 & 128 & relu & 1.1231$\times 10^{-3}$ & 2.1862$\times 10^{-4}$ \\ 
6 & 2 & 128 & 128 & -- & relu & 3.2495$\times 10^{-3}$ & 2.5605$\times 10^{-4}$ \\
7 & 3 & 32 & 32 & 128 & relu & 6.1855$\times 10^{-3}$ & 2.7093$\times 10^{-4}$ \\ 
8 & 3 & 32 & 128 & 32 & relu & 6.9244$\times 10^{-3}$ & 4.3660$\times 10^{-4}$ \\ 
9 & 2 & 128 & 128 & -- & relu & 8.5086$\times 10^{-4}$ & 4.4778$\times 10^{-4}$ \\ 
10 & 2 & 64 & 128 & -- & relu & 1.2916$\times 10^{-3}$ & 4.6174$\times 10^{-4}$ \\
\hline
\end{tabular}

%% file: table/MERGED_TABLE_1.tex
\begin{tabular}{@{}ccccccclllllllllllllllllllll@{}}
\toprule
OGLE ID  & $\lambda$   &        $\frac{\rm M}{ \rm M_{\sun}}$ & $\log \left( \frac{\rm L}{\rm L_{\sun}} \right)$ & $\teff$ (K)           & X (dex)               & Z (dex)               & P (d)                  &       $\rm m_{\rm obs}$  (mag) & $\rm M_{\rm ANN}$  (mag) & $\rm A_{\rm obs}$  (mag) & $\rm A_{\rm ANN}$   (mag)  &    $\rm R_{21_{\rm obs}}$ & $\rm R_{21_{\rm ANN}}$  & $\rm \phi_{21_{\rm obs}}$ & $\rm \phi_{21_{\rm ANN}}$ &   $R_{31_{\rm obs}}$    & $\rm R_{31_{\rm ANN}}$ & $\rm \phi_{31_{\rm obs}}$ & $\rm \phi_{31_{\rm ANN}}$ &    $R_{41_{\rm obs}}$ & $R_{41_{\rm ANN}}$ & $\rm \phi_{41_{\rm obs}}$ & $\rm \phi_{41_{\rm ANN}}$ &    $\rm R_{51_{\rm obs}}$ & $\rm R_{51_{\rm ANN}}$ & $\rm \phi_{51_{\rm obs}}$ & $\rm \phi_{51_{\rm ANN}}$ \\ \midrule

LMC-00008 &      I & 0.605$\pm$0.054 & 1.778$\pm$0.044 &   6490$\pm$110 & 0.753971 & 0.001029 & 0.7877152$\pm$3.1e-06 & 18.679$\pm$0.002 &   -0.155$\pm$0.001 &  0.462$\pm$0.076 &  0.663$\pm$0.234 & 0.385$\pm$0.024 &      0.419$\pm$0.005 &  3.257$\pm$0.078 &          3.267$\pm$0.015 & 0.253$\pm$0.024 &      0.119$\pm$0.005 &  0.334$\pm$0.114 &          6.173$\pm$0.042 & 0.046$\pm$0.023 &      0.045$\pm$0.005 &  3.834$\pm$0.506 &          2.089$\pm$0.105 & 0.024$\pm$0.024 &      0.024$\pm$0.005 &  1.518$\pm$0.946 &          3.454$\pm$0.199 \\
LMC-00010 &      I & 0.676$\pm$0.059 & 1.667$\pm$0.044 &   6380$\pm$110 & 0.754428 & 0.000572 & 0.5940743$\pm$1.6e-06 & 18.819$\pm$0.002 &    0.094$\pm$0.001 &  0.765$\pm$0.095 &  0.563$\pm$0.166 & 0.447$\pm$0.026 &       0.397$\pm$0.01 &  2.858$\pm$0.068 &          3.437$\pm$0.029 & 0.345$\pm$0.024 &      0.275$\pm$0.009 &  5.918$\pm$0.096 &          0.336$\pm$0.043 & 0.178$\pm$0.023 &      0.213$\pm$0.009 &   2.78$\pm$0.159 &          3.812$\pm$0.056 & 0.057$\pm$0.022 &       0.16$\pm$0.009 &  5.101$\pm$0.436 &           0.88$\pm$0.073 \\
LMC-00027 &      I & 0.714$\pm$0.056 & 1.828$\pm$0.045 &   6270$\pm$110 & 0.753610 & 0.001390 &   0.7938821$\pm$3e-06 & 18.507$\pm$0.002 &   -0.294$\pm$0.001 &  0.551$\pm$0.065 &  0.438$\pm$0.221 & 0.426$\pm$0.023 &      0.206$\pm$0.006 &  3.157$\pm$0.068 &          3.322$\pm$0.028 & 0.244$\pm$0.023 &       0.13$\pm$0.005 &  0.424$\pm$0.109 &          3.758$\pm$0.045 & 0.111$\pm$0.023 &      0.103$\pm$0.005 &   3.617$\pm$0.21 &          0.402$\pm$0.057 &  0.07$\pm$0.022 &      0.107$\pm$0.005 &  1.601$\pm$0.334 &          3.693$\pm$0.057 \\
LMC-00072 &      I & 0.693$\pm$0.055 & 1.698$\pm$0.045 &   6360$\pm$110 & 0.754293 & 0.000707 &   0.6261742$\pm$2e-06 & 18.886$\pm$0.002 &    0.027$\pm$0.001 &  0.558$\pm$0.068 &  0.554$\pm$0.169 & 0.415$\pm$0.021 &       0.373$\pm$0.01 &  2.923$\pm$0.062 &            3.4$\pm$0.031 &  0.28$\pm$0.021 &       0.275$\pm$0.01 &   6.155$\pm$0.09 &          0.187$\pm$0.044 &  0.144$\pm$0.02 &       0.198$\pm$0.01 &  3.379$\pm$0.156 &           3.687$\pm$0.06 & 0.042$\pm$0.019 &       0.14$\pm$0.009 &  0.744$\pm$0.481 &          0.771$\pm$0.082 \\
LMC-00082 &      I & 0.665$\pm$0.058 & 1.715$\pm$0.043 &   6690$\pm$110 & 0.754716 & 0.000284 &   0.5654446$\pm$8e-07 & 18.741$\pm$0.003 &    0.082$\pm$0.001 &  0.808$\pm$0.072 &  0.905$\pm$0.321 &  0.465$\pm$0.02 &      0.474$\pm$0.006 &  2.623$\pm$0.055 &          3.018$\pm$0.016 &  0.377$\pm$0.02 &      0.395$\pm$0.006 &  5.485$\pm$0.074 &          5.776$\pm$0.022 & 0.275$\pm$0.019 &      0.252$\pm$0.006 &  2.292$\pm$0.101 &          2.249$\pm$0.031 & 0.151$\pm$0.018 &      0.175$\pm$0.006 &  5.466$\pm$0.154 &          5.248$\pm$0.042 \\

\vdots & \vdots & \vdots         & \vdots                            & \vdots & \vdots & \vdots & \vdots & \vdots    & \vdots    & \vdots & \vdots    & \vdots     & \vdots  & \vdots  & \vdots & \vdots  & \vdots     & \vdots     & \vdots     & \vdots & \vdots     & \vdots     & \vdots     & \vdots & \vdots     & \vdots     \\
SMC-0001 &      I & 0.595$\pm$0.057 & 1.661$\pm$0.043 &   6640$\pm$110 & 0.754758 & 0.000242 &   0.5588145$\pm$9e-07 & 19.065$\pm$0.003 &    0.166$\pm$0.001 &  0.899$\pm$0.084 &  0.871$\pm$0.184 & 0.444$\pm$0.019 &      0.462$\pm$0.005 &  2.643$\pm$0.052 &          3.069$\pm$0.013 & 0.341$\pm$0.018 &      0.382$\pm$0.005 &  5.359$\pm$0.072 &          6.153$\pm$0.018 & 0.249$\pm$0.017 &      0.222$\pm$0.005 &  1.968$\pm$0.098 &          2.591$\pm$0.027 & 0.165$\pm$0.017 &       0.13$\pm$0.005 &   5.01$\pm$0.131 &          5.583$\pm$0.042 \\
SMC-0002 &      I & 0.581$\pm$0.054 & 1.604$\pm$0.043 &   6420$\pm$100 & 0.754461 & 0.000539 &  0.594794$\pm$1.9e-06 & 19.011$\pm$0.003 &    0.202$\pm$0.001 &  0.689$\pm$0.071 &  0.519$\pm$0.211 & 0.441$\pm$0.028 &      0.396$\pm$0.009 &  2.837$\pm$0.077 &          3.545$\pm$0.027 & 0.304$\pm$0.027 &      0.219$\pm$0.009 &  5.906$\pm$0.114 &          0.232$\pm$0.047 & 0.148$\pm$0.026 &      0.253$\pm$0.009 &  3.073$\pm$0.202 &          3.413$\pm$0.048 & 0.091$\pm$0.026 &      0.201$\pm$0.009 &  0.351$\pm$0.303 &           0.355$\pm$0.06 \\
SMC-0003 &      I &  0.648$\pm$0.05 & 1.723$\pm$0.044 &   6570$\pm$110 & 0.754293 & 0.000707 & 0.6506795$\pm$3.3e-06 & 19.158$\pm$0.002 &    0.019$\pm$0.001 &   0.911$\pm$0.24 &  0.797$\pm$0.183 & 0.323$\pm$0.027 &      0.444$\pm$0.005 &  2.938$\pm$0.099 &          3.186$\pm$0.015 & 0.179$\pm$0.027 &      0.344$\pm$0.005 &  6.173$\pm$0.168 &           6.072$\pm$0.02 & 0.052$\pm$0.026 &      0.176$\pm$0.005 &   3.502$\pm$0.52 &           2.67$\pm$0.034 & 0.018$\pm$0.027 &      0.086$\pm$0.005 &  0.495$\pm$1.459 &          5.644$\pm$0.062 \\
SMC-0008 &      I &  0.697$\pm$0.05 & 1.675$\pm$0.042 &   6270$\pm$100 & 0.754268 & 0.000732 & 0.6328767$\pm$2.1e-06 & 19.154$\pm$0.002 &    0.052$\pm$0.001 &  0.788$\pm$0.103 &    0.5$\pm$0.145 & 0.416$\pm$0.027 &      0.407$\pm$0.012 &  2.911$\pm$0.078 &          3.625$\pm$0.036 & 0.281$\pm$0.026 &      0.305$\pm$0.012 &  6.175$\pm$0.115 &           0.71$\pm$0.051 & 0.109$\pm$0.026 &      0.264$\pm$0.012 &  3.156$\pm$0.247 &          4.485$\pm$0.063 & 0.048$\pm$0.024 &      0.208$\pm$0.012 &  0.595$\pm$0.535 &          1.672$\pm$0.079 \\
SMC-0011 &      I & 0.669$\pm$0.051 & 1.697$\pm$0.042 &   6530$\pm$100 & 0.754651 & 0.000349 & 0.5957643$\pm$1.8e-06 &  19.17$\pm$0.003 &    0.075$\pm$0.001 &  0.962$\pm$0.134 &  0.738$\pm$0.189 & 0.359$\pm$0.029 &      0.419$\pm$0.005 &  2.649$\pm$0.094 &          3.171$\pm$0.015 & 0.268$\pm$0.029 &      0.321$\pm$0.005 &  5.682$\pm$0.129 &          6.065$\pm$0.021 & 0.149$\pm$0.028 &      0.194$\pm$0.005 &  2.501$\pm$0.209 &          2.779$\pm$0.032 & 0.099$\pm$0.028 &      0.109$\pm$0.005 &  5.522$\pm$0.303 &          5.897$\pm$0.052 \\

\vdots & \vdots & \vdots         & \vdots                            & \vdots & \vdots & \vdots & \vdots & \vdots    & \vdots    & \vdots & \vdots    & \vdots     & \vdots  & \vdots  & \vdots & \vdots  & \vdots     & \vdots     & \vdots     & \vdots   & \vdots     & \vdots     & \vdots     & \vdots    \\

LMC-00008 &      V & 0.605$\pm$0.054 & 1.778$\pm$0.044 &   6490$\pm$110 & 0.753971 & 0.001029 & 0.7877152$\pm$3.1e-06 & 19.419$\pm$0.013 &    0.384$\pm$0.001 &  0.583$\pm$0.079 &  0.828$\pm$0.377 & 0.377$\pm$0.089 &      0.339$\pm$0.005 &  2.798$\pm$0.256 &          2.752$\pm$0.018 & 0.179$\pm$0.071 &      0.193$\pm$0.005 &  5.868$\pm$0.557 &          5.965$\pm$0.031 & 0.105$\pm$0.065 &      0.018$\pm$0.005 &  3.083$\pm$0.883 &          1.997$\pm$0.287 & 0.096$\pm$0.072 &      0.032$\pm$0.005 &  0.003$\pm$0.906 &          0.847$\pm$0.163 \\
LMC-00010 &      V & 0.676$\pm$0.059 & 1.667$\pm$0.044 &   6380$\pm$110 & 0.754428 & 0.000572 & 0.5940743$\pm$1.6e-06 & 19.466$\pm$0.012 &    0.659$\pm$0.002 &  0.774$\pm$0.043 &    0.87$\pm$0.19 & 0.408$\pm$0.076 &       0.399$\pm$0.01 &   2.386$\pm$0.21 &          2.842$\pm$0.029 & 0.375$\pm$0.073 &       0.276$\pm$0.01 &  5.167$\pm$0.271 &          5.583$\pm$0.043 & 0.299$\pm$0.079 &       0.18$\pm$0.009 &   1.701$\pm$0.34 &          2.103$\pm$0.063 & 0.099$\pm$0.071 &      0.142$\pm$0.009 &  4.029$\pm$0.832 &          5.161$\pm$0.079 \\
LMC-00072 &      V & 0.693$\pm$0.055 & 1.698$\pm$0.045 &   6360$\pm$110 & 0.754293 & 0.000707 &   0.6261742$\pm$2e-06 & 19.563$\pm$0.011 &    0.588$\pm$0.002 &  0.672$\pm$0.039 &  0.833$\pm$0.181 & 0.424$\pm$0.067 &       0.388$\pm$0.01 &  2.429$\pm$0.175 &          2.867$\pm$0.031 & 0.252$\pm$0.056 &       0.257$\pm$0.01 &  5.347$\pm$0.279 &          5.576$\pm$0.047 &  0.17$\pm$0.061 &       0.175$\pm$0.01 &  2.014$\pm$0.368 &          2.165$\pm$0.067 & 0.094$\pm$0.064 &       0.132$\pm$0.01 &  5.655$\pm$0.634 &          5.127$\pm$0.086 \\
LMC-00079 &      V & 0.623$\pm$0.055 &  1.66$\pm$0.044 &   6590$\pm$110 & 0.754575 & 0.000425 & 0.5647915$\pm$1.4e-06 & 19.602$\pm$0.018 &    0.676$\pm$0.001 &  0.858$\pm$0.044 &   1.141$\pm$0.31 &  0.57$\pm$0.117 &      0.458$\pm$0.003 &  2.474$\pm$0.305 &          2.596$\pm$0.009 &   0.3$\pm$0.095 &      0.321$\pm$0.003 &  5.308$\pm$0.443 &          5.404$\pm$0.013 & 0.293$\pm$0.105 &       0.18$\pm$0.003 &  1.712$\pm$0.563 &          1.733$\pm$0.021 & 0.276$\pm$0.072 &      0.079$\pm$0.003 &  4.812$\pm$0.714 &          4.646$\pm$0.041 \\
LMC-00082 &      V & 0.665$\pm$0.058 & 1.715$\pm$0.043 &   6690$\pm$110 & 0.754716 & 0.000284 &   0.5654446$\pm$8e-07 & 19.314$\pm$0.019 &     0.58$\pm$0.002 &  1.001$\pm$0.035 &  1.343$\pm$0.306 & 0.393$\pm$0.077 &       0.49$\pm$0.006 &  2.278$\pm$0.258 &          2.625$\pm$0.015 & 0.276$\pm$0.085 &      0.328$\pm$0.005 &   4.938$\pm$0.34 &          5.176$\pm$0.022 & 0.201$\pm$0.071 &      0.234$\pm$0.005 &  1.496$\pm$0.528 &           1.339$\pm$0.03 & 0.123$\pm$0.071 &       0.15$\pm$0.005 &  3.158$\pm$0.786 &          4.025$\pm$0.042 \\

\vdots & \vdots & \vdots         & \vdots                            & \vdots & \vdots & \vdots & \vdots & \vdots    & \vdots    & \vdots & \vdots    & \vdots     & \vdots  & \vdots  & \vdots & \vdots  & \vdots     & \vdots     & \vdots     & \vdots  & \vdots     & \vdots     & \vdots     & \vdots & \vdots     & \vdots      \\
SMC-0001 &      V & 0.595$\pm$0.057 & 1.661$\pm$0.043 &   6640$\pm$110 & 0.754758 & 0.000242 &   0.5588145$\pm$9e-07 & 19.594$\pm$0.014 &    0.667$\pm$0.001 &  1.052$\pm$0.074 &   1.25$\pm$0.301 & 0.385$\pm$0.052 &      0.489$\pm$0.003 &   2.04$\pm$0.184 &          2.612$\pm$0.009 &  0.342$\pm$0.06 &      0.333$\pm$0.003 &  4.461$\pm$0.209 &          5.602$\pm$0.013 & 0.277$\pm$0.055 &      0.189$\pm$0.003 &  0.997$\pm$0.272 &           1.841$\pm$0.02 & 0.091$\pm$0.047 &      0.081$\pm$0.003 &    3.5$\pm$0.492 &          4.666$\pm$0.041 \\
SMC-0002 &      V & 0.581$\pm$0.054 & 1.604$\pm$0.043 &   6420$\pm$100 & 0.754461 & 0.000539 &  0.594794$\pm$1.9e-06 & 19.615$\pm$0.007 &    0.797$\pm$0.002 &  1.014$\pm$0.143 &  0.767$\pm$0.254 & 0.484$\pm$0.048 &      0.408$\pm$0.011 &   2.48$\pm$0.097 &          2.681$\pm$0.032 & 0.252$\pm$0.038 &        0.23$\pm$0.01 &  5.515$\pm$0.184 &          5.455$\pm$0.054 & 0.127$\pm$0.035 &       0.176$\pm$0.01 &   2.102$\pm$0.34 &          2.025$\pm$0.071 & 0.107$\pm$0.039 &        0.14$\pm$0.01 &  5.791$\pm$0.366 &          4.528$\pm$0.089 \\
SMC-0003 &      V &  0.648$\pm$0.05 & 1.723$\pm$0.044 &   6570$\pm$110 & 0.754293 & 0.000707 & 0.6506795$\pm$3.3e-06 & 19.767$\pm$0.004 &    0.539$\pm$0.001 &  0.794$\pm$0.152 &  1.041$\pm$0.334 &  0.331$\pm$0.03 &      0.447$\pm$0.005 &  2.568$\pm$0.117 &          2.612$\pm$0.013 &   0.22$\pm$0.03 &      0.309$\pm$0.005 &  5.338$\pm$0.168 &           5.413$\pm$0.02 & 0.077$\pm$0.028 &      0.149$\pm$0.005 &  2.339$\pm$0.413 &          1.503$\pm$0.035 & 0.046$\pm$0.028 &      0.053$\pm$0.004 &  0.287$\pm$0.617 &           4.14$\pm$0.087 \\
SMC-0008 &      V &  0.697$\pm$0.05 & 1.675$\pm$0.042 &   6270$\pm$100 & 0.754268 & 0.000732 & 0.6328767$\pm$2.1e-06 & 19.742$\pm$0.003 &    0.639$\pm$0.002 &  0.986$\pm$0.154 &  0.943$\pm$0.241 & 0.422$\pm$0.024 &      0.424$\pm$0.014 &  2.436$\pm$0.068 &           3.006$\pm$0.04 & 0.227$\pm$0.022 &      0.359$\pm$0.014 &  5.445$\pm$0.117 &          6.068$\pm$0.053 & 0.116$\pm$0.022 &      0.269$\pm$0.013 &  2.048$\pm$0.211 &          2.899$\pm$0.071 & 0.012$\pm$0.022 &      0.238$\pm$0.013 &  5.254$\pm$1.699 &          6.175$\pm$0.085 \\
SMC-0011 &      V & 0.669$\pm$0.051 & 1.697$\pm$0.042 &   6530$\pm$100 & 0.754651 & 0.000349 & 0.5957643$\pm$1.8e-06 &  19.81$\pm$0.008 &    0.605$\pm$0.001 &  1.054$\pm$0.102 &  1.043$\pm$0.251 & 0.436$\pm$0.037 &      0.426$\pm$0.005 &  2.048$\pm$0.115 &          2.647$\pm$0.013 & 0.215$\pm$0.036 &      0.284$\pm$0.004 &  4.605$\pm$0.201 &             5.3$\pm$0.02 & 0.102$\pm$0.036 &       0.17$\pm$0.004 &   0.84$\pm$0.357 &           1.685$\pm$0.03 &  0.05$\pm$0.035 &      0.081$\pm$0.004 &  2.629$\pm$0.718 &          4.565$\pm$0.057 \\

\vdots & \vdots & \vdots         & \vdots                            & \vdots & \vdots & \vdots & \vdots & \vdots    & \vdots    & \vdots & \vdots    & \vdots     & \vdots  & \vdots  & \vdots & \vdots  & \vdots     & \vdots     & \vdots     & \vdots & \vdots     & \vdots     & \vdots     & \vdots & \vdots     & \vdots      \\ \bottomrule
\end{tabular}

%% file: paper.bbl
\begin{thebibliography}{}
\makeatletter
\relax
\def\mn@urlcharsother{\let\do\@makeother \do\$\do\&\do\#\do\^\do\_\do\%\do\~}
\def\mn@doi{\begingroup\mn@urlcharsother \@ifnextchar [ {\mn@doi@}
  {\mn@doi@[]}}
\def\mn@doi@[#1]#2{\def\@tempa{#1}\ifx\@tempa\@empty \href
  {http://dx.doi.org/#2} {doi:#2}\else \href {http://dx.doi.org/#2} {#1}\fi
  \endgroup}
\def\mn@eprint#1#2{\mn@eprint@#1:#2::\@nil}
\def\mn@eprint@arXiv#1{\href {http://arxiv.org/abs/#1} {{\tt arXiv:#1}}}
\def\mn@eprint@dblp#1{\href {http://dblp.uni-trier.de/rec/bibtex/#1.xml}
  {dblp:#1}}
\def\mn@eprint@#1:#2:#3:#4\@nil{\def\@tempa {#1}\def\@tempb {#2}\def\@tempc
  {#3}\ifx \@tempc \@empty \let \@tempc \@tempb \let \@tempb \@tempa \fi \ifx
  \@tempb \@empty \def\@tempb {arXiv}\fi \@ifundefined
  {mn@eprint@\@tempb}{\@tempb:\@tempc}{\expandafter \expandafter \csname
  mn@eprint@\@tempb\endcsname \expandafter{\@tempc}}}

\bibitem[\protect\citeauthoryear{Alexander \& Ferguson}{Alexander \&
  Ferguson}{1994}]{alexander_low-temperature_1994}
Alexander D.~R.,  Ferguson J.~W.,  1994, \mn@doi [\apj] {10.1086/175039}, 437,
  879

\bibitem[\protect\citeauthoryear{Asplund, Grevesse  \& Sauval}{Asplund
  et~al.}{2005}]{asplund_solar_2005}
Asplund M.,  Grevesse N.,   Sauval A.~J.,  2005, in Barnes Thomas~G. I.,  Bash
  F.~N.,  eds,  Astronomical {Society} of the {Pacific} {Conference} {Series}
  Vol. 336, Cosmic {Abundances} as {Records} of {Stellar} {Evolution} and
  {Nucleosynthesis}. p.~25

\bibitem[\protect\citeauthoryear{{Astropy Collaboration} et~al.,}{{Astropy
  Collaboration} et~al.}{2013}]{astropy_collaboration_astropy_2013}
{Astropy Collaboration} et~al., 2013, \mn@doi [\aap]
  {10.1051/0004-6361/201322068}, 558, A33

\bibitem[\protect\citeauthoryear{{Astropy Collaboration} et~al.,}{{Astropy
  Collaboration} et~al.}{2018}]{astropy_collaboration_astropy_2018}
{Astropy Collaboration} et~al., 2018, \mn@doi [\aj] {10.3847/1538-3881/aabc4f},
  156, 123

\bibitem[\protect\citeauthoryear{Bailer-Jones, Rybizki, Fouesneau, Mantelet  \&
  Andrae}{Bailer-Jones et~al.}{2018}]{bailer-jones_estimating_2018}
Bailer-Jones C. A.~L.,  Rybizki J.,  Fouesneau M.,  Mantelet G.,   Andrae R.,
  2018, \mn@doi [\aj] {10.3847/1538-3881/aacb21}, 156, 58

\bibitem[\protect\citeauthoryear{Bailer-Jones, Rybizki, Fouesneau, Demleitner
  \& Andrae}{Bailer-Jones et~al.}{2021}]{bailer-jones_estimating_2021}
Bailer-Jones C. A.~L.,  Rybizki J.,  Fouesneau M.,  Demleitner M.,   Andrae R.,
   2021, \mn@doi [\aj] {10.3847/1538-3881/abd806}, 161, 147

\bibitem[\protect\citeauthoryear{Bellinger, Kanbur, Bhardwaj  \&
  Marconi}{Bellinger et~al.}{2020}]{bellinger_when_2020}
Bellinger E.~P.,  Kanbur S.~M.,  Bhardwaj A.,   Marconi M.,  2020, \mn@doi
  [\mnras] {10.1093/mnras/stz3292}, 491, 4752

\bibitem[\protect\citeauthoryear{Bergstra \& Bengio}{Bergstra \&
  Bengio}{2012}]{bergstra_random_2012}
Bergstra J.,  Bengio Y.,  2012, Journal of machine learning research, 13

\bibitem[\protect\citeauthoryear{Bhardwaj}{Bhardwaj}{2022}]{bhardwaj_rr_2022}
Bhardwaj A.,  2022, \mn@doi [Universe] {10.3390/universe8020122}, 8, 122

\bibitem[\protect\citeauthoryear{Bhardwaj, Kanbur, Singh, Macri  \&
  Ngeow}{Bhardwaj et~al.}{2015}]{bhardwaj_variation_2015}
Bhardwaj A.,  Kanbur S.~M.,  Singh H.~P.,  Macri L.~M.,   Ngeow C.-C.,  2015,
  \mn@doi [\mnras] {10.1093/mnras/stu2678}, 447, 3342

\bibitem[\protect\citeauthoryear{Bhardwaj, Macri, Rejkuba, Kanbur, Ngeow  \&
  Singh}{Bhardwaj et~al.}{2017a}]{bhardwaj_large_2017}
Bhardwaj A.,  Macri L.~M.,  Rejkuba M.,  Kanbur S.~M.,  Ngeow C.-C.,   Singh
  H.~P.,  2017a, \mn@doi [\aj] {10.3847/1538-3881/aa5e4f}, 153, 154

\bibitem[\protect\citeauthoryear{Bhardwaj, Kanbur, Marconi, Rejkuba, Singh  \&
  Ngeow}{Bhardwaj et~al.}{2017b}]{bhardwaj_comparative_2017}
Bhardwaj A.,  Kanbur S.~M.,  Marconi M.,  Rejkuba M.,  Singh H.~P.,   Ngeow
  C.-C.,  2017b, \mn@doi [\mnras] {10.1093/mnras/stw3256}, 466, 2805

\bibitem[\protect\citeauthoryear{Bhardwaj et~al.,}{Bhardwaj
  et~al.}{2021}]{bhardwaj_rr_2021}
Bhardwaj A.,  et~al., 2021, \mn@doi [\apj] {10.3847/1538-4357/abdf48}, 909, 200

\bibitem[\protect\citeauthoryear{Bono \& Stellingwerf}{Bono \&
  Stellingwerf}{1994}]{bono_pulsation_1994}
Bono G.,  Stellingwerf R.~F.,  1994, \mn@doi [\apjs] {10.1086/192054}, 93, 233

\bibitem[\protect\citeauthoryear{Bono, Caputo, Castellani  \& Marconi}{Bono
  et~al.}{1997}]{bono_nonlinear_1997}
Bono G.,  Caputo F.,  Castellani V.,   Marconi M.,  1997, \mn@doi [\aaps]
  {10.1051/aas:1997289}, 121, 327

\bibitem[\protect\citeauthoryear{Bono, Caputo  \& Marconi}{Bono
  et~al.}{1998}]{bono_theoretical_1998}
Bono G.,  Caputo F.,   Marconi M.,  1998, \mn@doi [\apjl] {10.1086/311270},
  497, L43

\bibitem[\protect\citeauthoryear{Bono, Marconi  \& Stellingwerf}{Bono
  et~al.}{1999}]{bono_classical_1999}
Bono G.,  Marconi M.,   Stellingwerf R.~F.,  1999, \mn@doi [\apjs]
  {10.1086/313207}, 122, 167

\bibitem[\protect\citeauthoryear{Bono, Castellani  \& Marconi}{Bono
  et~al.}{2000a}]{bono_classical_2000}
Bono G.,  Castellani V.,   Marconi M.,  2000a, \mn@doi [\apj] {10.1086/308263},
  529, 293

\bibitem[\protect\citeauthoryear{Bono, Castellani  \& Marconi}{Bono
  et~al.}{2000b}]{bono_rr_2000}
Bono G.,  Castellani V.,   Marconi M.,  2000b, \mn@doi [\apjl]
  {10.1086/312582}, 532, L129

\bibitem[\protect\citeauthoryear{Bono, Caputo, Castellani, Marconi  \&
  Storm}{Bono et~al.}{2001}]{bono_theoretical_2001}
Bono G.,  Caputo F.,  Castellani V.,  Marconi M.,   Storm J.,  2001, \mn@doi
  [\mnras] {10.1046/j.1365-8711.2001.04655.x}, 326, 1183

\bibitem[\protect\citeauthoryear{Caputo, Castellani, Degl'Innocenti, Fiorentino
   \& Marconi}{Caputo et~al.}{2004}]{caputo_bright_2004}
Caputo F.,  Castellani V.,  Degl'Innocenti S.,  Fiorentino G.,   Marconi M.,
  2004, \mn@doi [\aap] {10.1051/0004-6361:20040307}, 424, 927

\bibitem[\protect\citeauthoryear{Catelan, Pritzl  \& Smith}{Catelan
  et~al.}{2004}]{catelan_rr_2004}
Catelan M.,  Pritzl B.~J.,   Smith H.~A.,  2004, \mn@doi [\apjs]
  {10.1086/422916}, 154, 633

\bibitem[\protect\citeauthoryear{Clementini, Gratton, Bragaglia, Carretta,
  Fabrizio  \& Maio}{Clementini et~al.}{2003}]{clementini_distance_2003}
Clementini G.,  Gratton R.,  Bragaglia A.,  Carretta E.,  Fabrizio L.~D.,
  Maio M.,  2003, \mn@doi [The Astronomical Journal] {10.1086/367773}, 125,
  1309

\bibitem[\protect\citeauthoryear{Coppola et~al.,}{Coppola
  et~al.}{2011}]{coppola_distance_2011}
Coppola G.,  et~al., 2011, \mn@doi [\mnras] {10.1111/j.1365-2966.2011.19102.x},
  416, 1056

\bibitem[\protect\citeauthoryear{Cusano et~al.,}{Cusano
  et~al.}{2013}]{cusano_dwarf_2013}
Cusano F.,  et~al., 2013, \mn@doi [\apj] {10.1088/0004-637X/779/1/7}, 779, 7

\bibitem[\protect\citeauthoryear{Cybenko}{Cybenko}{1989}]{cybenko_approximation_1989}
Cybenko G.,  1989, \mn@doi [Mathematics of Control, Signals and Systems]
  {10.1007/BF02551274}, 2, 303

\bibitem[\protect\citeauthoryear{Das, Bhardwaj, Kanbur, Singh  \& Marconi}{Das
  et~al.}{2018}]{das_variation_2018}
Das S.,  Bhardwaj A.,  Kanbur S.~M.,  Singh H.~P.,   Marconi M.,  2018, \mn@doi
  [Monthly Notices of the Royal Astronomical Society] {10.1093/mnras/sty2358},
  481, 2000

\bibitem[\protect\citeauthoryear{Das et~al.,}{Das
  et~al.}{2020}]{das_stellar_2020}
Das S.,  et~al., 2020, \mn@doi [Monthly Notices of the Royal Astronomical
  Society] {10.1093/mnras/staa182}, 493, 29

\bibitem[\protect\citeauthoryear{De~Somma, Marconi, Molinaro, Cignoni, Musella
  \& Ripepi}{De~Somma et~al.}{2020}]{de_somma_extended_2020}
De~Somma G.,  Marconi M.,  Molinaro R.,  Cignoni M.,  Musella I.,   Ripepi V.,
  2020, \mn@doi [\apjs] {10.3847/1538-4365/ab7204}, 247, 30

\bibitem[\protect\citeauthoryear{De~Somma, Marconi, Molinaro, Ripepi, Leccia
  \& Musella}{De~Somma et~al.}{2022}]{de_somma_updated_2022}
De~Somma G.,  Marconi M.,  Molinaro R.,  Ripepi V.,  Leccia S.,   Musella I.,
  2022, \mn@doi [\apjs] {10.3847/1538-4365/ac7f3b}, 262, 25

\bibitem[\protect\citeauthoryear{Deb \& Singh}{Deb \&
  Singh}{2009}]{deb_light_2009}
Deb S.,  Singh H.~P.,  2009, \mn@doi [\aap] {10.1051/0004-6361/200912851}, 507,
  1729

\bibitem[\protect\citeauthoryear{Di~Criscienzo et~al.,}{Di~Criscienzo
  et~al.}{2011}]{di_criscienzo_new_2011}
Di~Criscienzo M.,  et~al., 2011, \mn@doi [\aj] {10.1088/0004-6256/141/3/81},
  141, 81

\bibitem[\protect\citeauthoryear{Drake et~al.,}{Drake
  et~al.}{2013}]{drake_probing_2013}
Drake A.~J.,  et~al., 2013, \mn@doi [\apj] {10.1088/0004-637X/763/1/32}, 763,
  32

\bibitem[\protect\citeauthoryear{Draper \& Smith}{Draper \&
  Smith}{1998}]{draper_applied_1998}
Draper N.~R.,  Smith H.,  1998, Applied regression analysis.
~ Vol. 326, John Wiley \& Sons

\bibitem[\protect\citeauthoryear{Elsken, Metzen  \& Hutter}{Elsken
  et~al.}{2019}]{elsken_neural_2019}
Elsken T.,  Metzen J.~H.,   Hutter F.,  2019, The Journal of Machine Learning
  Research, 20, 1997

\bibitem[\protect\citeauthoryear{Feuchtinger}{Feuchtinger}{1999}]{feuchtinger_nonlinear_1999}
Feuchtinger M.~U.,  1999, \mn@doi [\aaps] {10.1051/aas:1999462}, 136, 217

\bibitem[\protect\citeauthoryear{Fiorentino et~al.,}{Fiorentino
  et~al.}{2010}]{fiorentino_rr_2010}
Fiorentino G.,  et~al., 2010, \mn@doi [\apj] {10.1088/0004-637X/708/1/817},
  708, 817

\bibitem[\protect\citeauthoryear{Glantz \& Slinker}{Glantz \&
  Slinker}{2001}]{glantz_primer_2001}
Glantz S.~A.,  Slinker B.~K.,  2001, Primer of applied regression \& analysis
  of variance, ed.
McGraw-Hill, Inc., New York

\bibitem[\protect\citeauthoryear{Graczyk et~al.,}{Graczyk
  et~al.}{2014}]{graczyk_araucaria_2014}
Graczyk D.,  et~al., 2014, \mn@doi [\apj] {10.1088/0004-637X/780/1/59}, 780, 59

\bibitem[\protect\citeauthoryear{Guo, Yang, Wu, Wang  \& Liang}{Guo
  et~al.}{2008}]{guo_novel_2008}
Guo X.,  Yang J.,  Wu C.,  Wang C.,   Liang Y.,  2008, Neurocomputing, 71, 3211

\bibitem[\protect\citeauthoryear{Harris et~al.,}{Harris
  et~al.}{2020}]{harris_array_2020}
Harris C.~R.,  et~al., 2020, \mn@doi [Nature] {10.1038/s41586-020-2649-2}, 585,
  357

\bibitem[\protect\citeauthoryear{Haschke, Grebel  \& Duffau}{Haschke
  et~al.}{2011}]{haschke_new_2011}
Haschke R.,  Grebel E.~K.,   Duffau S.,  2011, \mn@doi [\aj]
  {10.1088/0004-6256/141/5/158}, 141, 158

\bibitem[\protect\citeauthoryear{Hornik}{Hornik}{1991}]{hornik_approximation_1991}
Hornik K.,  1991, \mn@doi [Neural Networks]
  {https://doi.org/10.1016/0893-6080(91)90009-T}, 4, 251

\bibitem[\protect\citeauthoryear{Hornik, Stinchcombe  \& White}{Hornik
  et~al.}{1989}]{hornik_multilayer_1989}
Hornik K.,  Stinchcombe M.,   White H.,  1989, \mn@doi [Neural Networks]
  {https://doi.org/10.1016/0893-6080(89)90020-8}, 2, 359

\bibitem[\protect\citeauthoryear{Howell et~al.,}{Howell
  et~al.}{2014}]{howell_k2_2014}
Howell S.~B.,  et~al., 2014, \mn@doi [\pasp] {10.1086/676406}, 126, 398

\bibitem[\protect\citeauthoryear{Hunter}{Hunter}{2007}]{hunter_matplotlib_2007}
Hunter J.~D.,  2007, \mn@doi [Computing in Science \& Engineering]
  {10.1109/MCSE.2007.55}, 9, 90

\bibitem[\protect\citeauthoryear{Iglesias \& Rogers}{Iglesias \&
  Rogers}{1996}]{iglesias_updated_1996}
Iglesias C.~A.,  Rogers F.~J.,  1996, \mn@doi [\apj] {10.1086/177381}, 464, 943

\bibitem[\protect\citeauthoryear{Jurcsik \& Kovacs}{Jurcsik \&
  Kovacs}{1996}]{jurcsik_determination_1996}
Jurcsik J.,  Kovacs G.,  1996, \aap, 312, 111

\bibitem[\protect\citeauthoryear{Jurcsik et~al.,}{Jurcsik
  et~al.}{2015}]{jurcsik_overtone_2015}
Jurcsik J.,  et~al., 2015, The Astrophysical Journal Supplement Series, 219, 25

\bibitem[\protect\citeauthoryear{Keller \& Wood}{Keller \&
  Wood}{2002}]{keller_large_2002}
Keller S.~C.,  Wood P.~R.,  2002, \mn@doi [\apj] {10.1086/342315}, 578, 144

\bibitem[\protect\citeauthoryear{Kingma \& Ba}{Kingma \&
  Ba}{2014}]{kingma_adam_2014}
Kingma D.~P.,  Ba J.,  2014, arXiv e-prints, p. arXiv:1412.6980

\bibitem[\protect\citeauthoryear{Kuehn et~al.,}{Kuehn
  et~al.}{2013}]{kuehn_rr_2013}
Kuehn C.~A.,  et~al., 2013, arXiv e-prints, p. arXiv:1310.0553

\bibitem[\protect\citeauthoryear{Kunder et~al.,}{Kunder
  et~al.}{2013}]{kunder_rr_2013}
Kunder A.,  et~al., 2013, \mn@doi [\aj] {10.1088/0004-6256/146/5/119}, 146, 119

\bibitem[\protect\citeauthoryear{Lenz \& Breger}{Lenz \&
  Breger}{2005}]{lenz_period04_2005}
Lenz P.,  Breger M.,  2005, \mn@doi [Communications in Asteroseismology]
  {10.1553/cia146s53}, 146, 53

\bibitem[\protect\citeauthoryear{Lomb}{Lomb}{1976}]{lomb_least-squares_1976}
Lomb N.~R.,  1976, \mn@doi [\apss] {10.1007/BF00648343}, 39, 447

\bibitem[\protect\citeauthoryear{Longmore, Fernley  \& Jameson}{Longmore
  et~al.}{1986}]{longmore_rr_1986}
Longmore A.~J.,  Fernley J.~A.,   Jameson R.~F.,  1986, \mn@doi [\mnras]
  {10.1093/mnras/220.2.279}, 220, 279

\bibitem[\protect\citeauthoryear{Luger, Agol, Kruse, Barnes, Becker,
  Foreman-Mackey  \& Deming}{Luger et~al.}{2016}]{luger_everest_2016}
Luger R.,  Agol E.,  Kruse E.,  Barnes R.,  Becker A.,  Foreman-Mackey D.,
  Deming D.,  2016, \mn@doi [\aj] {10.3847/0004-6256/152/4/100}, 152, 100

\bibitem[\protect\citeauthoryear{Luo et~al.,}{Luo
  et~al.}{2015}]{luo_first_2015}
Luo A.-L.,  et~al., 2015, \mn@doi [Research in Astronomy and Astrophysics]
  {10.1088/1674-4527/15/8/002}, 15, 1095

\bibitem[\protect\citeauthoryear{Madore}{Madore}{1982}]{madore_period-luminosity_1982}
Madore B.~F.,  1982, \mn@doi [\apj] {10.1086/159659}, 253, 575

\bibitem[\protect\citeauthoryear{{Marconi}}{{Marconi}}{2009}]{marconi_2009}
{Marconi} M.,  2009, in {Guzik} J.~A.,  {Bradley} P.~A.,  eds,  American
  Institute of Physics Conference Series Vol. 1170, Stellar Pulsation:
  Challenges for Theory and Observation. pp 223--234 (\mn@eprint {arXiv}
  {0909.0900}), \mn@doi{10.1063/1.3246450}

\bibitem[\protect\citeauthoryear{Marconi \& Clementini}{Marconi \&
  Clementini}{2005}]{marconi_pulsational_2005}
Marconi M.,  Clementini G.,  2005, \mn@doi [\aj] {10.1086/429525}, 129, 2257

\bibitem[\protect\citeauthoryear{Marconi \& Degl'Innocenti}{Marconi \&
  Degl'Innocenti}{2007}]{marconi_modeling_2007}
Marconi M.,  Degl'Innocenti S.,  2007, \mn@doi [\aap]
  {10.1051/0004-6361:20065840}, 474, 557

\bibitem[\protect\citeauthoryear{Marconi, Caputo, Di~Criscienzo  \&
  Castellani}{Marconi et~al.}{2003}]{marconi_rr_2003}
Marconi M.,  Caputo F.,  Di~Criscienzo M.,   Castellani M.,  2003, \mn@doi
  [\apj] {10.1086/377641}, 596, 299

\bibitem[\protect\citeauthoryear{Marconi, Bono, Caputo, Piersimoni,
  Pietrinferni  \& Stellingwerf}{Marconi et~al.}{2011}]{marconi_period_2011}
Marconi M.,  Bono G.,  Caputo F.,  Piersimoni A.~M.,  Pietrinferni A.,
  Stellingwerf R.~F.,  2011, \mn@doi [\apj] {10.1088/0004-637X/738/1/111}, 738,
  111

\bibitem[\protect\citeauthoryear{Marconi, Molinaro, Ripepi, Musella  \&
  Brocato}{Marconi et~al.}{2013}]{marconi_theoretical_2013}
Marconi M.,  Molinaro R.,  Ripepi V.,  Musella I.,   Brocato E.,  2013, \mn@doi
  [\mnras] {10.1093/mnras/sts197}, 428, 2185

\bibitem[\protect\citeauthoryear{Marconi et~al.,}{Marconi
  et~al.}{2015}]{marconi_new_2015}
Marconi M.,  et~al., 2015, \mn@doi [The Astrophysical Journal]
  {10.1088/0004-637x/808/1/50}, 808, 50

\bibitem[\protect\citeauthoryear{Marconi et~al.,}{Marconi
  et~al.}{2017}]{marconi_vmc_2017}
Marconi M.,  et~al., 2017, \mn@doi [\mnras] {10.1093/mnras/stw3289}, 466, 3206

\bibitem[\protect\citeauthoryear{Marconi, Bono, Pietrinferni, Braga, Castellani
   \& Stellingwerf}{Marconi et~al.}{2018}]{marconi_impact_2018}
Marconi M.,  Bono G.,  Pietrinferni A.,  Braga V.~F.,  Castellani M.,
  Stellingwerf R.~F.,  2018, \mn@doi [\apjl] {10.3847/2041-8213/aada17}, 864,
  L13

\bibitem[\protect\citeauthoryear{{Martín Abadi} et~al.,}{{Martín Abadi}
  et~al.}{2015}]{martin_abadi_tensorflow_2015}
{Martín Abadi} et~al., 2015, {TensorFlow}: {Large}-{Scale} {Machine}
  {Learning} on {Heterogeneous} {Systems}

\bibitem[\protect\citeauthoryear{McKinney}{McKinney}{2010}]{mckinney_data_2010}
McKinney W.,  2010, in Walt S. v.~d.,  Millman J.,  eds, Proceedings of the 9th
  {Python} in {Science} {Conference}. pp 56 -- 61,
  \mn@doi{10.25080/Majora-92bf1922-00a}

\bibitem[\protect\citeauthoryear{Moretti et~al.,}{Moretti
  et~al.}{2009}]{moretti_leo_2009}
Moretti M.~I.,  et~al., 2009, \mn@doi [\apjl] {10.1088/0004-637X/699/2/L125},
  699, L125

\bibitem[\protect\citeauthoryear{Mullen et~al.,}{Mullen
  et~al.}{2021}]{mullen_metallicity_2021}
Mullen J.~P.,  et~al., 2021, \mn@doi [\apj] {10.3847/1538-4357/abefd4}, 912,
  144

\bibitem[\protect\citeauthoryear{Muraveva et~al.,}{Muraveva
  et~al.}{2015}]{muraveva_new_2015}
Muraveva T.,  et~al., 2015, \mn@doi [\apj] {10.1088/0004-637X/807/2/127}, 807,
  127

\bibitem[\protect\citeauthoryear{Natale, Marconi  \& Bono}{Natale
  et~al.}{2008}]{natale_theoretical_2008}
Natale G.,  Marconi M.,   Bono G.,  2008, \mn@doi [\apjl] {10.1086/526518},
  674, L93

\bibitem[\protect\citeauthoryear{Nemec et~al.,}{Nemec
  et~al.}{2011}]{nemec_fourier_2011}
Nemec J.~M.,  et~al., 2011, \mn@doi [\mnras]
  {10.1111/j.1365-2966.2011.19317.x}, 417, 1022

\bibitem[\protect\citeauthoryear{Nemec, Cohen, Ripepi, Derekas, Moskalik,
  Sesar, Chadid  \& Bruntt}{Nemec et~al.}{2013}]{nemec_metal_2013}
Nemec J.~M.,  Cohen J.~G.,  Ripepi V.,  Derekas A.,  Moskalik P.,  Sesar B.,
  Chadid M.,   Bruntt H.,  2013, \mn@doi [\apj] {10.1088/0004-637X/773/2/181},
  773, 181

\bibitem[\protect\citeauthoryear{O'Malley, Bursztein, Long, Chollet, Jin,
  Invernizzi  \& {others}}{O'Malley et~al.}{2019}]{omalley_kerastuner_2019}
O'Malley T.,  Bursztein E.,  Long J.,  Chollet F.,  Jin H.,  Invernizzi L.,
  {others} 2019, {KerasTuner}, \url {https://github.com/keras-team/keras-tuner}

\bibitem[\protect\citeauthoryear{Paxton, Bildsten, Dotter, Herwig, Lesaffre  \&
  Timmes}{Paxton et~al.}{2011}]{paxton_modules_2011}
Paxton B.,  Bildsten L.,  Dotter A.,  Herwig F.,  Lesaffre P.,   Timmes F.,
  2011, \mn@doi [\apjs] {10.1088/0067-0049/192/1/3}, 192, 3

\bibitem[\protect\citeauthoryear{Paxton et~al.,}{Paxton
  et~al.}{2013}]{paxton_modules_2013}
Paxton B.,  et~al., 2013, \mn@doi [\apjs] {10.1088/0067-0049/208/1/4}, 208, 4

\bibitem[\protect\citeauthoryear{Paxton et~al.,}{Paxton
  et~al.}{2015}]{paxton_modules_2015}
Paxton B.,  et~al., 2015, \mn@doi [\apjs] {10.1088/0067-0049/220/1/15}, 220, 15

\bibitem[\protect\citeauthoryear{Paxton et~al.,}{Paxton
  et~al.}{2018}]{paxton_modules_2018}
Paxton B.,  et~al., 2018, \mn@doi [\apjs] {10.3847/1538-4365/aaa5a8}, 234, 34

\bibitem[\protect\citeauthoryear{Paxton et~al.,}{Paxton
  et~al.}{2019}]{paxton_modules_2019}
Paxton B.,  et~al., 2019, \mn@doi [\apjs] {10.3847/1538-4365/ab2241}, 243, 10

\bibitem[\protect\citeauthoryear{{Piersanti, L.}, {Straniero, O.}  \&
  {Cristallo, S.}}{{Piersanti, L.} et~al.}{2007}]{piersanti_l_method_2007}
{Piersanti, L.} {Straniero, O.}  {Cristallo, S.} 2007, \mn@doi [A\&A]
  {10.1051/0004-6361:20054505}, 462, 1051

\bibitem[\protect\citeauthoryear{Pietrinferni, Cassisi, Salaris  \&
  Castelli}{Pietrinferni et~al.}{2006}]{pietrinferni_large_2006}
Pietrinferni A.,  Cassisi S.,  Salaris M.,   Castelli F.,  2006, \mn@doi [\apj]
  {10.1086/501344}, 642, 797

\bibitem[\protect\citeauthoryear{Pietrukowicz et~al.,}{Pietrukowicz
  et~al.}{2015}]{pietrukowicz_deciphering_2015}
Pietrukowicz P.,  et~al., 2015, \mn@doi [\apj] {10.1088/0004-637X/811/2/113},
  811, 113

\bibitem[\protect\citeauthoryear{Pietrzyński et~al.,}{Pietrzyński
  et~al.}{2019}]{pietrzynski_distance_2019}
Pietrzyński G.,  et~al., 2019, \mn@doi [\nat] {10.1038/s41586-019-0999-4},
  567, 200

\bibitem[\protect\citeauthoryear{Ragosta et~al.,}{Ragosta
  et~al.}{2019}]{ragosta_vmc_2019}
Ragosta F.,  et~al., 2019, \mn@doi [\mnras] {10.1093/mnras/stz2881}, 490, 4975

\bibitem[\protect\citeauthoryear{Ruder}{Ruder}{2016}]{ruder_overview_2016}
Ruder S.,  2016, An overview of gradient descent optimization algorithms

\bibitem[\protect\citeauthoryear{{Rumelhart}, {Hinton}  \&
  {Williams}}{{Rumelhart} et~al.}{1986}]{Rumelhart86b}
{Rumelhart} D.~E.,  {Hinton} G.~E.,   {Williams} R.~J.,  1986, \mn@doi [nat]
  {10.1038/323533a0}, \href
  {https://ui.adsabs.harvard.edu/abs/1986Natur.323..533R} {323, 533}

\bibitem[\protect\citeauthoryear{Sandage, Katem  \& Sandage}{Sandage
  et~al.}{1981}]{sandage_oosterhoff_1981}
Sandage A.,  Katem B.,   Sandage M.,  1981, The Astrophysical Journal
  Supplement Series, 46, 41

\bibitem[\protect\citeauthoryear{Scargle}{Scargle}{1982}]{scargle_studies_1982}
Scargle J.~D.,  1982, \mn@doi [\apj] {10.1086/160554}, 263, 835

\bibitem[\protect\citeauthoryear{Schlafly \& Finkbeiner}{Schlafly \&
  Finkbeiner}{2011}]{schlafly_measuring_2011}
Schlafly E.~F.,  Finkbeiner D.~P.,  2011, \mn@doi [\apj]
  {10.1088/0004-637X/737/2/103}, 737, 103

\bibitem[\protect\citeauthoryear{Schlegel, Finkbeiner  \& Davis}{Schlegel
  et~al.}{1998}]{schlegel_maps_1998}
Schlegel D.~J.,  Finkbeiner D.~P.,   Davis M.,  1998, \mn@doi [\apj]
  {10.1086/305772}, 500, 525

\bibitem[\protect\citeauthoryear{Serenelli \& Basu}{Serenelli \&
  Basu}{2010}]{serenelli_determining_2010}
Serenelli A.~M.,  Basu S.,  2010, \mn@doi [The Astrophysical Journal]
  {10.1088/0004-637x/719/1/865}, 719, 865

\bibitem[\protect\citeauthoryear{Skowron et~al.,}{Skowron
  et~al.}{2016}]{skowron_ogle-ing_2016}
Skowron D.~M.,  et~al., 2016, \actaa, 66, 269

\bibitem[\protect\citeauthoryear{Smolec \& Moskalik}{Smolec \&
  Moskalik}{2008}]{smolec_convective_2008}
Smolec R.,  Moskalik P.,  2008, \actaa, 58, 193

\bibitem[\protect\citeauthoryear{Sollima, Borissova, Catelan, Smith, Minniti,
  Cacciari  \& Ferraro}{Sollima et~al.}{2006}]{sollima_new_2006}
Sollima A.,  Borissova J.,  Catelan M.,  Smith H.~A.,  Minniti D.,  Cacciari
  C.,   Ferraro F.~R.,  2006, \mn@doi [\apjl] {10.1086/503099}, 640, L43

\bibitem[\protect\citeauthoryear{Soszyński et~al.,}{Soszyński
  et~al.}{2009}]{soszynski_optical_2009}
Soszyński I.,  et~al., 2009, \actaa, 59, 1

\bibitem[\protect\citeauthoryear{Soszyński et~al.,}{Soszyński
  et~al.}{2016}]{soszynski_ogle_2016}
Soszyński I.,  et~al., 2016, \actaa, 66, 131

\bibitem[\protect\citeauthoryear{Soszyński et~al.,}{Soszyński
  et~al.}{2017}]{soszynski_ogle_2017}
Soszyński I.,  et~al., 2017, arXiv preprint arXiv:1712.01307

\bibitem[\protect\citeauthoryear{Soszyński et~al.,}{Soszyński
  et~al.}{2018}]{soszynski_ogle_2018}
Soszyński I.,  et~al., 2018, \mn@doi [\actaa] {10.32023/0001-5237/68.2.1}, 68,
  89

\bibitem[\protect\citeauthoryear{Steel, Torrie  \& {others}}{Steel
  et~al.}{1960}]{steel_principles_1960}
Steel R. G.~D.,  Torrie J.~H.,   {others} 1960, Principles and procedures of
  statistics.

\bibitem[\protect\citeauthoryear{Stellingwerf}{Stellingwerf}{1982}]{stellingwerf_convection_1982}
Stellingwerf R.~F.,  1982, \mn@doi [\apj] {10.1086/160426}, 262, 339

\bibitem[\protect\citeauthoryear{Stellingwerf}{Stellingwerf}{1984}]{stellingwerf_convection_1984}
Stellingwerf R.~F.,  1984, \mn@doi [\apj] {10.1086/162454}, 284, 712

\bibitem[\protect\citeauthoryear{Vivas \& Zinn}{Vivas \&
  Zinn}{2006}]{vivas_quest_2006}
Vivas A.~K.,  Zinn R.,  2006, \mn@doi [\aj] {10.1086/505200}, 132, 714

\bibitem[\protect\citeauthoryear{Wang, Fu, Zong, Wang  \& Zhang}{Wang
  et~al.}{2021}]{wang_asteroseismology_2021}
Wang J.,  Fu J.-N.,  Zong W.,  Wang J.,   Zhang B.,  2021, \mn@doi [\mnras]
  {10.1093/mnras/stab1705}, 506, 6117

\bibitem[\protect\citeauthoryear{Waskom}{Waskom}{2021}]{waskom_seaborn_2021}
Waskom M.~L.,  2021, \mn@doi [Journal of Open Source Software]
  {10.21105/joss.03021}, 6, 3021

\bibitem[\protect\citeauthoryear{Zinn, Horowitz, Vivas, Baltay, Ellman,
  Hadjiyska, Rabinowitz  \& Miller}{Zinn et~al.}{2014}]{zinn_silla_2014}
Zinn R.,  Horowitz B.,  Vivas A.~K.,  Baltay C.,  Ellman N.,  Hadjiyska E.,
  Rabinowitz D.,   Miller L.,  2014, \mn@doi [\apj]
  {10.1088/0004-637X/781/1/22}, 781, 22

\bibitem[\protect\citeauthoryear{team}{team}{2020}]{team_pandas-devpandas_2020}
team T. p.~d.,  2020, pandas-dev/pandas: {Pandas},
  \mn@doi{10.5281/zenodo.3509134}, \url
  {https://doi.org/10.5281/zenodo.3509134}

\bibitem[\protect\citeauthoryear{{van Albada} \& {Baker}}{{van Albada} \&
  {Baker}}{1971}]{van_albada_1971}
{van Albada} T.~S.,  {Baker} N.,  1971, \mn@doi [\apj] {10.1086/151144}, \href
  {https://ui.adsabs.harvard.edu/abs/1971ApJ...169..311V} {169, 311}

\makeatother
\end{thebibliography}
